\documentclass[useAMS,usenatbib]{mn2e}

\usepackage{subfigure}
\usepackage{epsfig}
\usepackage{natbib}
\usepackage{amsmath,amsfonts,amssymb}

\usepackage{threeparttable}
\usepackage{txfonts}

\title[Sunyaev-Zel'dovich observation of Abell 2146]
{Sunyaev-Zel'dovich observation of the Bullet-like cluster Abell 2146 with
  the Arcminute Microkelvin Imager \thanks{We request that any
    reference to this paper cites `AMI Consortium: Rodr\'{i}guez-Gonz\'{a}lvez et al.2010}}

\author[AMI Consortium: Rodr\'{i}guez-Gonz\'{a}lvez et al.]
 {AMI Consortium:
 Carmen~Rodr\'{i}guez-Gonz\'{a}lvez$^1$\thanks{Email:cr384@cam.ac.uk}
 Malak~Olamaie,$^1$\thanks{Email:mo323@cam.ac.uk}
 \newauthor
  Matthew~L.~Davies,$^1$
  Andy~C.~Fabian,$^2$
  Farhan~Feroz,$^1$
 \newauthor
  Thomas~M.~O.~Franzen,$^1$
  Keith~J.~B.~Grainge,$^{1,3}$
  Michael~P.~Hobson,$^1$
  \newauthor
  Natasha~Hurley-Walker,$^1$
  Anthony~N.~Lasenby,$^{1,3}$
  Guy~G.~Pooley,$^1$
 \newauthor
  Helen~R.~Russell,$^2$
  Jeremy~S.~Sanders,$^2$
  Richard~D.~E.~Saunders,$^{1,3}$
 \newauthor
  Anna~M.~M.~Scaife,$^{4}$
  Michel~P.~Schammel,$^1$
  Paul~F.~Scott,$^1$
  \newauthor
  Timothy~W.~Shimwell,$^1$
  David~J.~Titterington,$^1$
  Elizabeth~M.~Waldram,$^1$
  \newauthor
  and Jonathan~T.~L.~Zwart$^5$\\
 $^1$ Astrophysics Group, Cavendish Laboratory,
      19 J.~J.~Thomson Avenue, Cambridge CB3 0HE \\
 $^2$ Institute of Astronomy, Madingley Road, Cambridge CB3 0HA \\
 $^3$ Kavli Institute for Cosmology Cambridge,
      Madingley Road, Cambridge, CB3 0HA \\
 $^4$ Dublin Institute for Advanced Studies, 31 Fitzwilliam Place,
      Dublin 2, Ireland \\
 $^5$ Columbia Astrophysics Laboratory, Columbia University, 550 West
      120th Street, New York, NY 10027, USA}
\date{Accepted ????. Received ????}
\pubyear{2010}

\begin{document}
\maketitle
\label{firstpage}

\begin{abstract}
We present $13.9-18.2$-GHz observations of the Sunyaev-Zel'dovich (SZ)
effect towards Abell 2146 using the Arcminute
Microkelvin Imager (AMI).
The cluster is detected with a peak signal-to-noise ratio of $13\sigma$ in
the radio source subtracted map from  $9$ hours of data. Comparison of the
SZ image with the X-ray image from \cite{russell09} suggests that both have extended
regions which lie approximately perpendicular to one another, with their 
emission peaks significantly displaced. These features indicate
non-uniformities in the distributions of the gas temperature and pressure, and
suggest complex dynamics indicative of a cluster merger.
We use a fast, Bayesian cluster analysis to
explore the high-dimensional parameter space of the
cluster-plus-sources model to obtain robust cluster parameter
estimates in the presence of radio point sources, receiver
noise and primordial CMB anisotropy; despite the substantial radio
emission from the direction of Abell 2146, the probability of SZ + CMB
primordial structure + radio sources + receiver noise to CMB + radio
sources + receiver noise is $3\times 10^{6}:1$. We compare the results from
three different cluster models. Our preferred model exploits the
observation that the gas fractions do not appear to vary greatly
between clusters. Given the relative masses of the
two merging systems in Abell 2146, the mean gas temperature can be deduced from the
virial theorem (assuming all of the kinetic energy is in the form of
internal gas energy) without being affected significantly by the
merger event, provided the primary cluster was virialized before the
merger. In this model we fit a simple spherical
isothermal $\beta$-model to our data, despite the inadequacy of this
model for a merging system like Abell 2146, and assume the cluster follows
the mass-temperature relation of a virialized, singular, isothermal
sphere. We note that this model avoids inferring large-scale cluster parameters
internal to $r_{200}$ under the widely used
assumption of hydrostatic equilibrium. We find that at
$r_{200}$ the average total mass $M_{\rm{T}}= \left(4.1 \pm
0.5 \right) \times 10^{14} h^{-1}\rm{M}_{\odot}$ and the mean gas
temperature $T=4.5 \pm 0.5$ keV.
\end{abstract}


\section{Introduction}\label{Introduction}

Galaxy clusters are the largest collapsed structures
known to exist in the Universe. The masses of rich clusters can reach
$\approx10^{15}h^{-1}\textrm{{M}}_{\odot}$ and the more distant ones, from
around $z>0.2$, subtend several arcminutes on the sky due to the slow variation of the angular diameter
distance with redshift. As a result, clusters
are powerful tracers of structure formation and evolution on scales
of the order of a few megaparsecs. According to the standard $\Lambda$
Cold Dark Matter ($\Lambda\textrm{CDM}$) model, galaxy clusters form
via hierarchical interactions of smaller
subsystems. During merger, these subclusters collide at
relative velocities of thousands of km$\rm{s}^{-1}$ and can release gravitational binding
energies of up to $\sim10^{57}\ \textrm{J}$, which can lead to
shocks in the intracluster medium (IM). These conditions make cluster
mergers ideal places to study
the dynamics of matter under extreme
conditions. The three assumed main components comprising the cluster, namely
galaxies, hot ionized gas and dark matter, exhibit very different behaviours
during subcluster mergers. The hot intergalactic gas is heated
and compressed by the hydrodynamical shocks produced during the
passage of the subcluster through the core of the primary,
whereas the dark matter and galaxies are collisionless (see e.g. \cite{markevitch2007}). As a result, the gas is slowed
down by ram pressure and is displaced from the dark matter and the galaxies. Later, when the subcluster reaches regions of lower gas
density in the primary cluster, the ram pressure drops
sharply. Without as much ram pressure, the gas pressure and subcluster
gravity cause some of the subcluster gas, which had been lagging behind
the subcluster's dark matter centre, to `slingshot' past it. This gas
is then left unbound from the subcluster and free to
expand adiabatically \citep{hallman2004}.

Abell 2146 is a cluster at $z=0.23$ consisting of two merging subclusters. The smaller
subcluster passed through the centre of the larger subcluster some
$0.1-0.3$ Gyrs ago producing shock fronts which have been
detected by Chandra \citep{russell09}. These shock fronts are 
unusual features which only show at a specific stage in the cluster
merger, before the shock reaches the outer, low-surface-brightness
regions, and at angles on the sky
plane which usually prevents the projection from hiding the density
edge. Therefore, it is not surprising that shock fronts with Mach
numbers significantly greater than one have only been detected in two other clusters:
1E0657-56 \citep{markevitch2002} --the ``Bullet cluster''-- and A520 \citep{markevitch2006}.
Unlike A520, the Bullet cluster and Abell 2146 appear to be at an early stage of the
merger event, where the cluster dynamics are simpler and the separation of
the hot gas and the dark matter components is clearer.

The thermal Sunyaev-Zeldovich (SZ) effect provides an independent
way of exploring the physics of the intracluster gas and examining typical
cluster parameters such as core radius and gas mass. When Cosmic
Microwave Background (CMB) photons traverse a rich galaxy cluster some
will be inverse-Compton scattered by the random thermal motion of the electrons
in the intracluster gas \citep{sunyaev1970, birkinshaw1999}. Unlike
X-ray surface brightness, SZ surface brightness is independent of redshift and is
therefore well suited for the study of galaxy clusters at any redshift.
It is also less sensitive than X-ray measurements to small-scale clumping and
the complex dynamics associated with the cluster core.

In this paper we present $16$-GHz SZ effect images of Abell 2146 using AMI. In
Section \ref{telescope} we discuss the telescope, while details
of the observations and the
reduction pipeline are given in Section \ref{obsprep}. In Section
\ref{bay} Bayesian inference is introduced. Section \ref{Physical-Model-and}
describes our analysis methodology while in Section \ref{res} and
Section \ref{dis} we present the 
results and discuss their significance. We present our conclusions
in Section \ref{conc}.

Throughout the paper we assume a concordance $\Lambda$CDM cosmology
with $\Omega_{\rm{m},0}=0.3$, $\Omega_{\Lambda,0}=0.7$, $\Omega_k=0$,
$\Omega_b=0.041$, $w_0=-1$, $w_a=0$, $\sigma_8=0.8$ and
$H_0=100$\:$\textrm{km\,s}^{-1}\textrm{Mpc}^{-1}$. Relevant parameters
are given in terms of the dimensionless Hubble parameter
$h=H_0/100$ km\,s$^{-1}$\,Mpc$^{-1}$, except where otherwise
stated. We also refer to $h_{X}=H_0/X$ km\,s$^{-1}$\,Mpc$^{-1}$. All
coordinates are at epoch J2000.

\section{The Telescope} \label{telescope}
AMI comprises two arrays: the Small Array (SA) which
consists of ten $3.7$-m diameter antennas, and
the Large Array (LA) with eight $13$-m antennas, located
at Lord's Bridge, Cambridge \citep{zwart2008}. The higher resolution and flux sensitivity of the LA allows
contaminating radio sources to be dealt with. These sources can then be
subtracted from the SA maps. A summary of the technical details of AMI is given in Table \ref{tab:tech}. Further details on the telescope can be
found in \cite{zwart2008}. 

\begin{table}
\caption{AMI technical summary}
\label{tab:tech}
\centering
\begin{centering}
\begin{tabular}{|c|c|c|}
\hline
  & { SA} & {LA}\tabularnewline
\hline
\hline 
{Antenna diameter} & {$3.7$\:m} & {$12.8$\:m}\tabularnewline
\hline 
{Number of antennas} & {10} & { 8}\tabularnewline
\hline 
{ Baseline lengths (current)} & {$5-20$\:m} & {$18-110$\:m}\tabularnewline
\hline 
{ Primary beam at $15.7$\:GHz} & {$20^{'}.1$} & { $5^{'}.5$}\tabularnewline
\hline 
{ Synthesized beam} & { $\approx3^{'}$} & { $\approx30^{''}$}\tabularnewline
\hline 
{ Flux sensitivity} & { $3\textrm{0 mJy s}^{-1/2}$} & { $3\textrm{ mJy s}^{-1/2}$}\tabularnewline
\hline 
{ Observing frequency} & { $13.9-18.2$GHz} & { $13.9-18.2$GHz}\tabularnewline
\hline 
{ Bandwidth} & { $4.3$\:GHz} & { $4.3$\:GHz}\tabularnewline
\hline 
{ Number of channels} & { 6} & { 6}\tabularnewline
\hline 
{ Channel bandwidth} & { $0.72$\:GHz} & { $0.72$\:GHz}\tabularnewline
\hline
\end{tabular}
\end{centering}

\end{table}

\section{ Observations and DATA Reduction} \label{obsprep}
\begin{table}
\centering
\caption{Assumed I+Q flux densities of 3C286 and 3C48, and errors
on flux measurements in each frequency channel, over the commonly-used
AMI SA bandwidth.}
\label{tab:fluxden}
\centering{}
\begin{tabular}{|c|c|c|c|c|}
\hline 
{\ Channel} & {\ $\nu/\textrm{GHz}$} & {\ $S^{\textrm{3C286}}/\textrm{Jy}$} & {\ $S^{\textrm{3C48}}/\textrm{Jy}$} & {\ $\sigma_{S}$}\tabularnewline
\hline
\hline 
{\ 3} & {\ $14.2$} & {\ $3.61$} & {\ $1.73$} & {\ $6.5\%$}\tabularnewline
\hline 
{\ 4} & {\ $15.0$} & {\ $3.49$} & {\ $1.65$} & {\ $5.0\%$}\tabularnewline
\hline 
{\ 5} & {\ $15.7$} & {\ $3.37$} & {\ $1.57$} & {\ $4.0\%$}\tabularnewline
\hline 
{\ 6} & {\ $16.4$} & {\ $3.26$} & {\ $1.49$} & {\ $3.5\%$}\tabularnewline
\hline 
{\ 7} & {\ $17.1$} & {\ $3.16$} & {\ $1.43$} & {\ $4.0\%$}\tabularnewline
\hline 
{\ 8} & {\ $17.9$} & {\ $3.06$} & {\ $1.37$} & {\ $7.0\%$}\tabularnewline
\hline
\end{tabular}
\end{table}

Observations of Abell 2146 were made by the SA and LA between November
$2009$ and March $2010$, yielding approximately $9$ hours of good
quality SA data; approximately the same amount of data suffered from artifacts
and was discarded. Data reduction was
perfomed using {\sc{reduce}}, a local software tool developed for the
Very Small Array (VSA) \citep{watson2003} and AMI (see
e.g. \cite{zwart2008} for further details). This package is designed
to apply path delay corrections and a series of algorithms tailored
to remove automatically bad data points arising from interference,
shadowing, hardware and other errors. We apply amplitude clips at a
$3\sigma$ level. Periods where the data has been
contaminated by interference are excised. These interference signals are identified
as persistent high amplitude signals in the lag domain, which appear in all the lag
channels. The system temperature
is monitored by a modulated noise signal sent to the front-end of each
antenna and synchronously detected at the end of each
intermediate-frequency channel and 
is used in {\sc{reduce}} to correct the amplitude scale
on an antenna basis. If the system temperature falls below $10\%$ of
the nominal value of an antenna the associated datapoints are
removed. For further details on the AMI reduction pipeline see
\cite{hurley2009}. Additional
manual flagging of remaining bad data points is done to ensure the
quality of the data. The correlator data are then Fourier
transformed into the frequency domain and stored on disk in FITS format.

Flux calibration was performed using short observations of primary calibrators, either 3C48 or 3C286.
The flux densities for 3C48 and 3C286, see Table \ref{tab:fluxden}, are in agreement
with \cite{baars1977} at 16-GHz. Since \cite{baars1977} measure $\mathrm{I}$, as opposed to AMI
which measures $\mathrm{I + Q}$, the flux densities were corrected
by interpolating from VLA $5$-, $8$- and $22$-\:GHz observations. Previous tests have shown
this calibration to be accurate to better than 5 per cent
\citep{scaife2009}. 
The phase is calibrated using interleaved calibrators
selected from the Jodrell Bank VLA Survey \citep{patnaik1992, browne1998,wilkinson1998} based
on their proximity and flux density. The phase calibrators used
for the observations of Abell 2146 were J1642$+$6856 for the SA and J1623$+$6624
for the LA. These phase calibrators were interleaved approximately
every hour for the SA and every ten minutes for the LA.

\subsection{Source subtraction }\label{Source-Subtraction}
Contamination from radio point sources at $\approx15$ GHz can
significantly obscure the SZ signal and must therefore be taken into account
in any SZ effect analyses at these frequencies. The higher resolution and flux sensitivity of the LA
is exploited to determine the position of the sources in the
SA maps accurately in a short amount of time. Local maxima on the continuum LA
maps above $4\sigma_n$, where $\sigma_{n}$ is the corresponding value
in Janskys per beam at that pixel in the
noise map, are identified as LA detected sources using AMI-developed source extraction
software (see \cite{franzen2010}).
Out of these LA-detected sources only those which appear within 0.1 of the
SA power primary beam having an apparent flux above $4\sigma_n$ on the SA map are
included in the source model. 

Every source in the source model is parameterized by a
position, a spectral index and a flux density whose priors are based on the LA measurements. The
source model is analysed by {\sc{McAdam}} (Monte Carlo
Astronomical Detection and Measurement), a Bayesian analysis package
for cluster detection and parameter extraction developed by \cite{marshall2003}
and adapted for AMI by \cite{feroz2009}, which fits a
probability distribution to the source flux densities at the positions given
by the LA. The source flux densities are fitted by  {\sc{McAdam}}  to allow for
possible intercalibration difference between the two AMI arrays and
for source variability. The
mean source flux-density values are then used to subtract the
sources from the SA map. 

\section{Bayesian Analysis of Clusters }\label{bay}

\subsection{Bayesian inference} \label{bayesian}
The cluster analysis software implemented in this paper \citep{marshall2003} is based on Bayesian inference. This robust methodology constrains a
set of parameters, $\boldsymbol{\Theta}$, given a model or hypothesis,
$H$ and the corresponding data, $\boldsymbol{D}$, using Bayes' theorem:

\begin{equation}
\textrm{Pr}(\boldsymbol{\Theta}\vert\boldsymbol{D},H)\equiv\frac{\textrm{Pr}(\boldsymbol{D}\vert\boldsymbol{\Theta},H)\ \textrm{Pr}(\boldsymbol{\Theta}\vert H)}{\textrm{\textrm{Pr}}(\boldsymbol{D}\vert H)}.\end{equation}
Here $\textrm{Pr}(\boldsymbol{\Theta}\vert\boldsymbol{D},H) \equiv P(\boldsymbol{\Theta})$
is the posterior probability distribution of the parameters, $\textrm{Pr}(\boldsymbol{D}\vert\boldsymbol{\Theta},H)\equiv\mathcal{L}(\boldsymbol{\Theta})$
is the likelihood, $\textrm{Pr}(\boldsymbol{\Theta}\vert H)\equiv\pi(\boldsymbol{\Theta})$
is the prior probability distribution and $\textrm{Pr}(\boldsymbol{D}\vert H)\equiv\mathcal{Z}$ the Bayesian evidence. If chosen wisely, incorporating
the prior knowledge into the analysis reduces the amount of parameter
space to be sampled and allows meaningful model selection.
 Bayesian inference can serve as a tool for two main purposes:
\begin{enumerate}
\item {Parameter estimation---In this case, the evidence factor can be neglected
since it is independent of the model parameters, $\boldsymbol{\Theta}$.
Sampling techniques can then be used
to explore the unnormalized posterior distributions. One obtains a set of samples from the
parameter space distributed according to the posterior. Constraints
on individual parameters can then be obtained by marginalising over
the other parameters.}
\item {Model selection---The evidence is crucial for ranking
models for the data. It is defined as the factor required for normalising the posterior over $\boldsymbol{\Theta}$:\begin{equation}
\mathcal{Z}=\int\mathcal{L}(\boldsymbol{\Theta})\pi(\boldsymbol{\Theta})d^{D}\boldsymbol{\Theta},\label{eq:evidence}\end{equation}
where $D$ is the dimensionality of the parameter space. This factor
represents an average of the likelihood over the prior and will therefore
favour models with high likelihood values throughout the entirety
of parameter space. This satisfies Occam's razor which states that
the evidence will be larger for simple models with compact parameter
spaces than for more complex ones, unless the latter fit the data
significantly better. Deciding which of two models, $H_{0}$ and $H_{1}$,
best fits the data can be done by computing the ratio\begin{equation}
 \frac{\textrm{Pr}(H_{1}\vert\boldsymbol{D})}{\textrm{Pr}(H_{0}\vert\boldsymbol{D})}=\frac{\textrm{Pr}(\boldsymbol{D}\vert
  H_{1})\textrm{Pr}(H_{1})}{\textrm{Pr}(\boldsymbol{D}\vert
  H_{0})\textrm{Pr}(H_{0})}=\frac{\mathcal{Z}_{1}}{\mathcal{Z}_{0}}\frac{\textrm{Pr}(H_{1})}{\textrm{Pr}(H_{0})},\label{eq:modelsel} \end{equation}
where $\frac{\textrm{Pr}(H_{1})}{\textrm{Pr}(H_{0})}$ is the prior
probability ratio set before any conclusions have been drawn from
the dataset.} 
\end{enumerate}

\subsection{Nested sampling}

Nested sampling is a Monte Carlo method introduced by \cite{skilling2004} which focuses on the efficient calculation
of evidences and generates posterior distributions as a
by-product. \cite{feroz2008} and \cite{bridges2009} have developed this sampling
framework and implemented the {\sc{MultiNest}} algorithm.
This algorithm can sample from posterior distributions
where multiple modes and/or large (curving) degeneracies are
present. This robust technique has reduced by a factor of $\approx100$ the computational costs
incurred during Bayesian parameter estimation and model selection. For
this reason the analysis in this paper is based on this technique.

\section{Physical Model and Assumptions }\label{Physical-Model-and}

\subsection{Interferometric data model}

An interferometer, like AMI, operating at a frequency,
$\nu$, measure samples from the complex visibility plane $\widetilde{I}_{\nu}(\boldsymbol{u})$.
These are given by a weighted Fourier transform of the surface brightness,
$I_{\nu}(\boldsymbol{x})$:\begin{equation}
\widetilde{I}_{\nu}(\boldsymbol{u})=\int A_{\nu}(\boldsymbol{x})I_{\nu}(\boldsymbol{x})\textrm{exp}(2\pi i\boldsymbol{u}\cdot\boldsymbol{x})\textrm{d}^{2}\boldsymbol{x},\end{equation}
where $\boldsymbol{x}$ is the position relative to the phase centre,
$A_{\nu}(\boldsymbol{x})$ is the (power) primary beam of the antennas
at an observing frequency, $\nu$ (normalized to unity at its peak)
and $\boldsymbol{u}$ is the baseline vector in units of wavelength.
In our model we assume the measured visibilities can be defined as
\begin{equation}
V_{\nu}(\boldsymbol{u}_{i})=\widetilde{I}_{\nu}(\boldsymbol{u}_{i})+N_{\nu}(\boldsymbol{u}_{i}),\end{equation}
where $\widetilde{I}_{\nu}(\boldsymbol{u})$ is the signal component,
which contains contributions from the cluster SZ effect signal and identified
radio point sources and $N_{\nu}(\boldsymbol{u}_{i})$ is 
a generalized noise component that includes signals from unresolved
point sources, primordial CMB anisotropies and instrumental noise.

\subsection{Cluster models }\label{clusparam}
In order to calculate the contribution of the cluster SZ signal to the
visibility data the Comptonization parameter of the cluster, $y(s)$,
across the sky must
be determined (see \cite{feroz2009} for further details). This
parameter is the integral of the gas pressure along the line of sight $l$ through the
cluster:
\begin{equation}
y(s)=\frac{\sigma_{T}}{m_{\rm{{e}}}c^{2}}\intop_{-\infty}^{\infty}n_{\rm{{e}}}k_{\rm{{B}}}T\textrm{d}l\propto\int_{-r_{\textrm{lim}}}^{+r_{\textrm{lim}}}\rho_{\rm{{g}}}T\textrm{d}l,
\label{eq:compton} 
\end{equation}
where $\sigma_{\rm{T}}$ is the Thomson scattering cross-section, $n_{\rm{e}}$ is the
electron number density, which is derived from equation (\ref{eq:ne}), $m_{\rm{e}}$ is the electron mass, $c$ is the speed
of light and $k_{\rm{B}}$ is the Boltzmann constant. $s=\theta D_{\theta}$
is the deprojected radius such that $r^{2}=s^{2}+l^{2}$ and
$D_{\theta}$ is the angular diameter distance to the cluster which can be
calculated for clusters at redshifts, $z$, using 
\begin{equation}
D_{\theta} =\frac{c\int^z_0 H^{-1}\left(z'\right)\textrm{d}z'}{\left(1+z\right)}. 
\end{equation}
We set $r_{\rm{lim}}$ in equation (\ref{eq:compton}) to $20h^{-1}\textrm{Mpc}$---this result has been tested and
shown to be large enough even for small values of $\beta$
\citep{marshall2003}. 

The cluster geometry, as well as two linearly independent functions of
its temperature and density profiles, must be specified to compute the Comptonization parameter. For the cluster geometry we have chosen a spherical
cluster model as a first approximation. The temperature profile
is assumed to be constant throughout the cluster. An isothermal
$\beta$-model is assumed for the cluster gas density, $\rho_g$
\citep{cavaliere1978}:
\begin{equation}
\rho_{\rm{g}}(r)=\frac{\rho_{\rm{g}}(0)}{\left[1+\left(\frac{r}{r_{\rm{c}}}\right)^{2}\right]^{\frac{3\beta}{2}}},\label{eq:beta}
\end{equation}
where 
\begin{equation}
\rho_{\rm{g}}(r)=\mu_{\rm{e}}n_{\rm{e}}(r), 
\label{eq:ne}
\end{equation}
$\mu_{\rm{e}}=1.14m_{\rm{p}}$ is the gas mass per electron and $m_{\rm{p}}$ is the proton mass. The core radius, $r_{\rm{c}}$, gives the
density profile a flat top at low $\frac{r}{r_{\rm{c}}}$ and
$\rho_{\rm{{g}}}$ has a logarithmic slope of
$3\beta$ at large $\frac{r}{r_{\rm{c}}}$.

Parameter estimates can depend on the way the cluster model is
parameterized. We
examine the impact of different physical assumptions by
presenting the parameter estimates for Abell 2146 obtained using three
different cluster parameterizations (or `models').
Modelled sources for all three models are characterised by three
parameters: position, flux density
and spectral index. The corresponding priors for these parameters are
given in Section \ref{par:sprior}. The parameterizations of the sources
and the source priors are the same in all three models, unlike the
cluster parameterizations which do change between models. The mean
values fitted by our {\sc{McAdam}} software to both the source and cluster
sampling parameters will, however, vary for each
cluster model. We proceed to describe our three cluster parameterizations and their results.

Tables \ref{tab:modassump} and \ref{tab:modderi} indicate which
parameters are derived in each model and the assumptions made in each case.
A summary of the sampling parameters for each model together with their priors is given in Table \ref{tab:Summary-of-cluster}.

\subsubsection{Cluster model 1}\label{model1}

Our first model, henceforth M1, is based on traditional methods for the analysis
of SZ and X-ray data.  
The sampling parameters for M1 are:
\begin{itemize}
\item { $\left(x_{\rm{c}},y_{\rm{c}}\right)$---the position of the cluster
centroid on the sky.}
\item {$T$---the temperature of the cluster gas, which is assumed to be
  uniform.}
\item { $\beta$---defines the outer logarithmic slope of the $\beta$-profile.}
\item { $r_{\rm{c}}$---gives the density profile a flat top at low $r$}
\item { $M_{\rm{g}}(r_{200})$---the gas mass inside a radius, $r_{200}$,
which is the radius at which the average total density is $200$
times $\rho_{\rm{crit}}$, the critical density for closure of the Universe.}
\item { $z$---the cluster redshift.}
\end{itemize}

In applying this cluster model to Abell 2146 both $z$ and $T$ are assumed to be known,
which is equivalent to assigning them delta-function priors (see Table \ref{tab:Summary-of-cluster}) .

The derived parameters for M1 are:
\begin{itemize}
\item {$r_{X}$---the radius at which the average total density is $X$ times $\rho_{\rm{crit}}$.}
\item {$M_{\rm{T}}(r_{X})$---the total cluster mass within the radius $r_{X}$.}
\item {$M_{g}(r_{X})$---the cluster gas mass within the radius $r_{X}$.}
\item {$f_{\rm{g}}(r_{X})$---the cluster average gas fraction within the radius $r_{X}$.}
\item {$\rho_g(0)$---the central gas density.}
\item {$y(0)$---the central Comptonization parameter.}
\end{itemize}

In this model, the cluster gas is assumed to be in hydrostatic equilibrium with the
total gravitational potential of the cluster, $\Phi$, which is
dominated by dark matter. As a result, the gravitational potential must satisfy
\begin{equation}
\frac{\textrm{d}\Phi}{\textrm{d}r}=-\frac{1}{\rho_{\rm{g}}}\frac{\textrm{d}p}{\textrm{d}r}.
\end{equation}
This equation can be simplified if the cluster gas consists purely of
ideal gas with a uniform temperature,
$T$, to give
\begin{equation}
\frac{\textrm{d log} \rho_{\rm{g}}}{\textrm{dlog}r}=-\frac{\rm{G}\mu}{\rm{k}_{\rm{B}} T}\frac{M_{\rm{T}}\left(r\right)}{r},\label{eq:hydro}
\end{equation}
where $\mu$ is the mass per particle,
$\mu\approx0.6m_{\rm{p}}\approx\frac{0.6}{1.14}\mu_{\rm{e}}$ \citep[see][]{marshall2003}.
Expressions for the total mass of the cluster, $M_{\rm{T}}(r_X)$, can
be obtained for spherical symmetry:
\begin{equation}
M_{\rm{T}}(r_X)=\frac{4\pi}{3}r_X^3 X\rho_{crit},
\label{eq:mtsphsym}
\end{equation}
or by integrating the isothermal $\beta$-model for the density profile in (\ref{eq:hydro}),
\begin{equation}
M_{\rm{T}}(r_X) = \frac{r_X^3}{r^2_{\rm{c}}+r_X^2}\frac{3\beta \rm{k}_{\rm{B}} \it{T}}{\rm{G}\mu}.
\label{eq:mtbeta}
\end{equation}
Combining equations (\ref{eq:mtsphsym}) and (\ref{eq:mtbeta}) leads to
an expression for $r_X$,
\begin{equation}
r_X=\sqrt{\frac{9\beta \rm{k_B} \it{T}}{4\pi \mu \rm{G} X \rho_{crit}}-r^2_{\rm{c}}}.
\label{eq:rm1}
\end{equation}
The total mass of the cluster within a certain radius, $M_{\rm{T}}(r_X)$,
is subsequently determined by substituting $r_X$ into equation (\ref{eq:mtsphsym}). Once $M_{\rm{T}}(r_X)$ and $M_{\rm{g}}(r_X)$ are
known, the gas fraction, $f_{\rm{g}}(r_X)$, can be computed using the relation
\begin{equation}
f_{\rm{g}}(r_X) = \frac{M_{\rm{g}}(r_X)}{M_{\rm{T}}(r_X)}.
\label{eq:fg}
\end{equation}
We consider values for $X=200$ and $500$. For $X=500$, $M_{\rm{g}}(r_X)$ is not a sampling parameter but is calculated using the expression 
\begin{equation}
M_{\rm{g}}(r_X)= \rho_{\rm{g}}(0) \int^{r_X}_0 \frac{4\pi r'^2}{\left[1+\frac{r'^2}{r_{\rm{c}}^2} \right]^{\frac{3\beta}{2}}}\textrm{d}r',
\label{eq:mgxgt2}
\end{equation}
Also, $\rho_{\rm{g}}(0)$, in equation (\ref{eq:beta}), can be recovered by
numerically integrating the gas density profile up to $r_{200}$,
equation (\ref{eq:mgxgt2}), and setting the result equal to $M_{\rm{g}}(r_{200})$.

\subsubsection{Cluster model 2} \label{model2}

Our second model, M2, has the same sampling parameters as M1 with the exception of $T$, which becomes
 a derived parameter, and $f_{\rm{g}}(r_{200})$, which becomes a
 sampling parameter.  Sampling from $f_{\rm{g}}(r_{200})$ and $M_{\rm{g}}(r_{200})$ allows $M_{\rm{T}}(r_{200})$ to be calculated using
equation (\ref{eq:fg}). $r_{200}$ can then be computed simply by rearranging equation (\ref{eq:mtsphsym}).
The temperature of the cluster gas can be obtained by combining equations
(\ref{eq:hydro}) and (\ref{eq:beta}) to yield
\begin{equation}
T=\frac{\rm{G} \mu}{3 \rm{k}_{\rm{B}} \beta}\frac{M_{\rm{T}}(r_{200})(r^2_{\rm{c}}+r_{200}^2)}{r_{200}^3},
\label{eq:temphydro}
\end{equation}
which is based upon the assumption that the cluster is in hydrostatic equilibrium
and described well by a $\beta$-profile. 
The derived parameters at $r_{500}$ are calculated in
the same way as in M1; once $M_{\rm{g}}(r_{500})$ is obtained from
equation (\ref{eq:mgxgt2}) and $r_{500}$ from equation (\ref{eq:rm1}),
$M_{\rm{T}}(r_{500})$ is calculated by assuming the cluster is
spherical, equation (\ref{eq:mtsphsym}). $f_{\rm{g}}(r_{500})$ can
then be recovered using the relation in (\ref{eq:fg}) .

\subsubsection{Cluster model 3}\label{model3}
In the third model, M3, the sampling and derived parameters are the same as
in M2. The only difference between M2 and M3 is the way $T$ is calculated. M3 uses an M-T relation to
derive $T$ which allows $T$ to be obtained without relying on the
cluster being in hydrostatic equilibrium, a necessary assumption in M2. Moreover, at
$r_{200}$, all the other cluster parameter estimates of M3 are free
from the assumption of hydrostatic equilibrium. However, this
assumption needs to be made to obtain cluster parameters at $r_{500}$
(see Section \ref{model2}).

If the cluster is assumed to be virialized and to contain a small
 amount of unseen energy density in the form of turbulence, bulk
 motions or magnetic fields, the average cluster gas temperature, $T$,
 can be obtained using the mass-temperature (M-T) relation for a
 singular, isothermal sphere (SIS) based on the virial theorem,

\begin{align}
\rm{k}_{\rm{B}}\it{T} &= \frac{\rm{G}\mu \it{M}_{\rm{T}}}{2r_{200}} \\
&= \frac{\rm{G}\mu}{2\left(\frac{3}{4\pi\left(200\rho_{\rm{crit}}\right)}\right)^{1/3}}M_{\rm{T}}^{2/3} \\
&= 8.2 \textrm{keV}\left(\frac{M_{\rm{T}}}{10^{15}h^{-1}\rm{M}_{\odot}}\right)^{2/3}\left(\frac{H(z)}{H_{0}}\right)^{2/3},
\label{eq:virtemp}
\end{align}
where H is the Hubble parameter. In our cluster model we use the well-behaved
 $\beta$-profile, equation (\ref{eq:beta}), rather than the
 SIS density profile which is singular at $r=0$. This different choice
 for the density profile will introduce a factor to the
 M-T relation in equation (\ref{eq:virtemp}). From cluster simulations we find that this
 factor varies between $ 0.7-1.2$.

\subsubsection{M-T relation and hydrostatic equilibrium}

The results obtained from running  {\sc{McAdam}} with three different models are useful for assessing the validity of some of the assumptions made in each
model. Traditional models tend to assume clusters are isothermal,
spherical, virialized and in hydrostatic equilibrium. All of these assumptions are
particularly inappropriate for cluster mergers like Abell 2146. The first two
assumptions are made in the three models presented in this paper to
simplify the cluster model; but note that the spherical assumption is
not bad here because our SZ measurements are sensitive to the larger
scales of the cluster.

M2 also assumes hydrostatic equilibrium to obtain an
estimate for $T$. After the gravitational collapse of a cluster, the hot
gas in the ICM tends to reach equilibrium when the force exerted by the thermal
pressure gradient of the ICM balances that from the cluster's own gravitational
force. An underlying assumption is that the gas pressure is provided
entirely by thermal pressure. In reality, there are many non-thermal
sources of pressure support present in most clusters such as turbulent
gas motions which can provide $\approx10-20\%$ of the total pressure support even in relaxed clusters
\citep{shuecker2004, rasia2006}. In the case of Abell 2146, a complex merging
system with two detected shocks propagating at $\approx1900$ and $2200$$
\textrm{kms}^{-1}$ \citep{russell09},
there is significant non-thermal pressure support provided by bulk
motions in the ICM. 

Relating radius, temperature and total mass via the virial theorem in practice also assumes that the kinetic
energy is in the form of internal energy of the particles, as evidenced
by the SZ signal, so that turbulent motions,
bulk motions and everything else are ignored. But this use of the virial theorem has an
advantage over hydrostatic equilibirum in the case of Abell 2146 since our knowledge of the mass ratio of
the two merging systems enables us to set a limit on the degree to
which the use of the M-T relation,  $T\propto M_{\rm{T}}^{\frac{2}{3}}$,
biases our temperature estimate. 

\cite{russell09} find
the fractional mass of the merging cluster to be between $25$ and $33$
percent, in which case the average temperature of the merging system
will be $\approx 10\%$ higher when all the gas mass of the subcluster has
merged with that of the primary cluster than prior to the start of the
merger event. Therefore, provided the primary cluster was virialized
pre-merger, our estimate for $T$ using the M-T relation in equation
(\ref{eq:virtemp}) is little affected by the merger.

\begin{table}
\caption{Summary of the derived parameters for each cluster model.}
 \label{tab:modderi}
 \centering
 \begin{centering}
 \begin{tabular}{|c|c|}\hline
{Derived Parameter} & {Model} \\ \hline \hline
{$r_{200}$ \& $r_{500}/h^{-1}$ Mpc} & {All} \\ \hline
{$M_{\rm{T}}(r_{200})$ \& $M_{\rm{T}}(r_{500})/h^{-1}\textrm{M}_\odot$} & {All} \\ \hline
{$M_{\rm{g}}(r_{500})/h^{-2}\textrm{M}_\odot$} & {All} \\ \hline
{$y$} & {All} \\ \hline 
{$n_{\rm{e}}$} & {All} \\ \hline 
{$T$ keV} & {M2, M3} \\ \hline 
{$f_{\rm{g}}(r_{200})/h^{-1}$ } & {M1} \\ \hline 
{$f_{\rm{g}}(r_{500})/h^{-1}$} & {All} \\ \hline 
\end{tabular}
\end{centering}
\end{table}

\begin{table*}

\centering
\begin{centering}
\newcommand*{\temp}[1]{\multicolumn{1}{|r|}{#1}}
\caption{Summary of the main assumptions made in the calculation of
  the derived parameters for each model. H stands for hydrostatic
  equilibrium, M-T for the mass-temperature relation given in equation (\ref{eq:virtemp})
  , B for isothermal $\beta$-profile,
  S for spherical geometry and N/A means not applicable, since that
  parameter is a sampling parameter for that particular model.}
\label{tab:modassump}
\begin{tabular}{|c|c|c|c|}\hline
{} & \multicolumn{3}{|c|}{Model Assumptions}\\  \hline \hline
{Derived Parameter} & {Model 1}  & {Model 2} & {Model 3} \\
\hline 
{$r_{200} /h^{-1}$ Mpc} & {H, S, B;
  Eq. \ref{eq:rm1}} & {S; Eq.\ref{eq:mtsphsym}} & {S;
  Eq.\ref{eq:mtsphsym}} \\ \hline
{$M_{\rm{T}}(r_{200})$}  & {S; Eq.\ref{eq:mtsphsym}} &
{Eq.\ref{eq:fg}} & {Eq.\ref{eq:fg}}  \\ \hline
{$f_{\rm{g}}(r_{200})/h^{-1}$ }  & {Eq.\ref{eq:fg}} & {N/A} & {N/A} \\
\hline 
{$M_{\rm{g}}(r_{500})/h^{-2}\textrm{M}_\odot$} & {S, B; Eq.\ref{eq:mgxgt2}}  & {S, B;
  Eq.\ref{eq:mgxgt2}}  & {S, B; Eq.\ref{eq:mgxgt2}}  \\ \hline
{$r_{500}/h^{-1}$ Mpc} & {H, S, B;
  Eq.\ref{eq:rm1}} & {H, S, B; Eq.\ref{eq:rm1}} & {H, S, B; Eq.\ref{eq:rm1}} \\ \hline
{$M_{\rm{T}}(r_{500})$}  & {S; Eq.\ref{eq:mtsphsym}} &
{S; Eq.\ref{eq:mtsphsym}} & {S; Eq.\ref{eq:mtsphsym}}  \\ \hline
{$f_{\rm{g}}(r_{500})/h^{-1}$ }  & {Eq.\ref{eq:fg}} & {Eq.\ref{eq:fg}} & {Eq.\ref{eq:fg}} \\
\hline 
{$T$ keV} &  {N/A} & {H; Eq.\ref{eq:temphydro}} & {M-T; Eq.\ref{eq:virtemp}} \\ \hline 
\end{tabular}
\end{centering}
\end{table*}

\subsection{Priors}
\subsubsection{Cluster priors}

For simplicity the priors are assumed to be separable. The priors used in the analysis of Abell 2146 are given in
Table \ref{tab:Summary-of-cluster}. 

We note that, although the prior on $M_{\rm{g}}(r_{200})$ assumes
the cluster produces a non-zero SZ effect, it is wide enough that our
results will not be biased. In fact, our posterior distributions for
$M_{\rm{g}}(r_{500})$ peak at $M_{\rm{g}}(r_{500}) >3 \times
10^{13}/h^{-2}\rm{M}_{\odot}$ and have fallen to zero by $M_{\rm{g}}(r_{500})>2 \times
10^{13}/h^{-2}\rm{M}_{\odot}$, while our prior for
$M_{\rm{g}}(r_{200})$ extends down to $1 \times
10^{13}/h^{-2}\rm{M}_{\odot}$

\begin{table*}
\centering

\caption{Summary of the priors for the sampling parameters in each
  model.}

\label{tab:Summary-of-cluster}
\begin{centering}
{ }\begin{tabular}{|c|c|c|c|c|}
\hline 
{ Parameter} & {Models} & { Prior Type} & { Values} & { Origin}\tabularnewline
\hline
\hline 
{{} $x_{\rm{c}},y_{\rm{c}} {''}$} & { all} & { gaussian at
  $\boldsymbol{x}_{\textrm{X-ray}},\sigma=60^{''}$} &
{$15^{\rm{h}}56^{\rm{m}}07^{\rm{s}},\ +66^{\circ}21\farcm 35\farcs$}
& { \cite{ebeling2000}}\tabularnewline
\hline 
{ $\beta$} &  { all} &  {uniform } & { $0.3-2.5$} & {\cite{marshall2003}}\tabularnewline
\hline 
{ $M_{\rm{g}}(r_{200})/h^{-2}\rm{M}_\odot$} & { all} & { uniform in log} & { $10^{13}-10^{15}$} & { physically reasonable}\tabularnewline
\hline 
{ $r_{\rm{c}}/h^{-1}\textrm{kpc}$} & { all} & { uniform} & { $10-1000$} & { physically reasonable}\tabularnewline
\hline 
{ $z$} & { all} & { delta} & { $0.23$} & {\cite{ebeling2000}}\tabularnewline
\hline 
{ $f_{\rm{g}}(r_{200})/h^{-1}$} & { M2, M3} & { gaussian, $\sigma=0.016$} & { $0.12$} & { \cite{larson2010}}\tabularnewline
\hline 
{ $T$ keV} & { M1} & { delta} & {6.7} & {\cite{russell09}}\tabularnewline
\hline
\end{tabular}
\end{centering}
\end{table*}

The prior on the gas mass fraction was set to a Gaussian centered at
the WMAP7 best-fit value, $f_{\rm{g}}=0.12h^{-1}$, with  $\sigma=0.016h^{-1}$. This result was
obtained from WMAP7 estimates of $\Omega_{\rm{m}}=0.266\pm 0.029$,
$\Omega_{\rm{b}}=0.0449\pm 0.0028$ and $h=0.710\pm 0.025$ using the relation 
$f_{\rm{b}}=\frac{\Omega_{\rm{b}}}{\Omega_{\rm{m}}}$, where
$f_{\rm{b}}$ is the universal baryon fraction \citep{larson2010}. The
prior on $f_{\rm{g}}$ can be based on $f_{\rm{b}}$ since $f_{\rm{g}}$ in clusters at large radii approaches $f_{\rm{b}}$.
The prior on the position of the cluster was a Gaussian with $\sigma=60''$ centered at the X-ray centroid.

\subsubsection{Source priors \label{par:sprior}}

As with the cluster priors, the source priors are assumed
to be separable, such that\[
\pi\left(\boldsymbol{\Theta}_{S}\right)=\pi\left(x_{\rm{s}}\right)\pi\left(y_{\rm{s}}\right)\pi\left(S_{0}\right)\pi\left(\alpha\right).\]

$\pi\left(x_{\rm{s}}\right)$ and $\pi\left(y_{\rm{s}}\right)$  are given delta
priors at the source position found from the high-resolution LA
maps. The flux-density priors for modelled sources on the other
hand are chosen to be Gaussians centered on the flux-density value given by the
LA with $\sigma\approx40\%$ of the LA source flux. 
Tight constraints on the flux-density priors are best avoided
due to inter-array calibration differences and source variability.
The channel flux densities taken from the LA data are used to calculate an
estimate for the spectral index of each source. The spectral index
prior is then set as a Gaussian centered at the predicted LA value
with a width $\sigma=1$. 

\section{Results }\label{res}
\subsection{Maps and evidences}

\begin{figure}
\centering
\includegraphics[width=7.5cm,height=
    7.5cm,clip=,angle=0.]{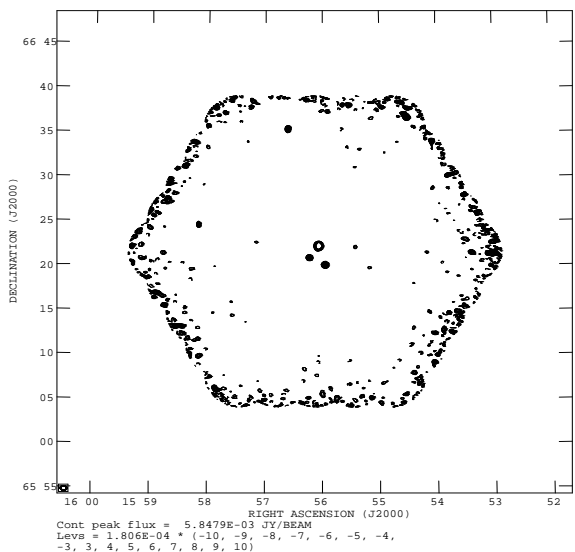}
\caption{LA contour map.}
\label{fig:LA}
\end{figure}

Fifteen sources were detected above $4\sigma_n$ on the
LA map, Fig. \ref{fig:LA}. {\sc{McAdam}} was used to determine the
flux densities and spectral indices of these sources in the SA data. The standard  {\sc{AIPS}} tasks were used to
{\sc{clean}} the images with a single {\sc{clean}} box. No primary
beam correction has been applied to the AMI maps
presented in this paper such that the thermal noise, $\sigma_n$, is
constant throughout the map. The task {\sc{imean}} was applied to the data to determine the noise level
on the maps. Contours increasing linearly in units
of $\sigma_n$ were used to produce all the contour maps. The
half-power contour of the synthesized beam for each map is shown at
the bottom left of each map.

\begin{figure}
\centering
\includegraphics[width=7.5cm,height= 7.5cm,clip=,angle=0.]{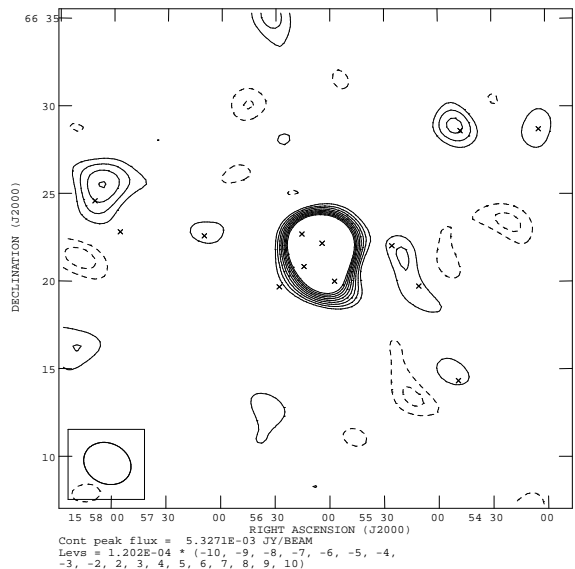}
\caption{SA map before source subtraction. The crosses indicate the position of the sources detected on the LA map.}
\label{fig:SAmbef}
\end{figure}

\begin{figure}
\centering
\includegraphics[width=7.5cm,height=7.5cm,clip=,angle=0.]{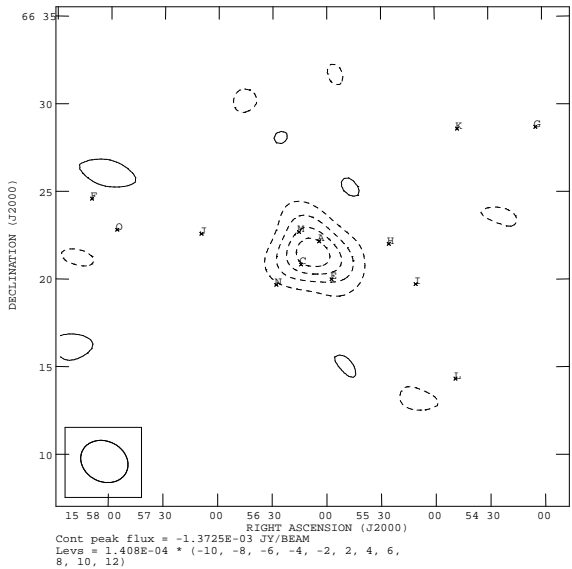}
\caption{SA contour map after source subtraction. The
letters represent the position of the sources detected on the LA
map.}
\label{fig:SAmaf}
\end{figure}

Further analysis was undertaken in the visibility plane taking into
account receiver noise, radio sources, and contributions from primary
CMB imprints. Figs \ref{fig:SAmbef} and \ref{fig:SAmaf} show the SA
maps of Abell 2146 before and after radio source
subtraction. The source subtraction was performed at the LA source
position using the mean flux-density estimates given by the {\sc{McAdam}}
results of M3,
Table \ref{tab:McAsourceFlux}. Sources with a high signal-to-noise ratio
and close to the pointing centre tend to have good agreement between
flux densities measured by the LA and those obtained by {\sc{McAdam}}. Possible
reasons for source flux-density discrepancies between the arrays, in
particular for the remaining sources, include: a poorer fit of the
Gaussian modelled primary beam at large $uv$-distances from the
pointing centre, loss of signal due to the white light fringes
 falling off the end of the correlator, time and bandwidth smearing,
 correlator artifacts, source variability and, some
 sources with low signal-to-noise ratios detected on the LA, might appear as noise features on the SA.

It should be noted that, since a single flux-density value is used for
subtracting the modelled sources in the map-plane, the radio source subtracted map
does not reflect the uncertainty in the {\sc{McAdam}} derived flux-density
estimates \footnote{Note that, unlike for the radio
  source subtracted maps, when obtaining estimates for the cluster
parameters the whole probability distribution for the source flux density is taken into account, such that a larger
uncertainty in the source flux densities will lead to wider
distributions in the cluster parameters.}. Nevertheless, flux-density
estimates  given by {\sc{McAdam}} have
been tested in \cite{feroz2009} and shown to be reliable. Fig. \ref{fig:massfluxrel} shows that there is no degeneracy between the
flux density fitted for source A in Fig. \ref{fig:SAmaf} and the
fitted values for $M_{\rm{g}}(r_{200})$, the  cluster gas mass within $r_{200}$
 The detection of Abell 2146 in the AMI data is confirmed by comparing the
evidence obtained by running {\sc{McAdam}} with a model including SZ +
CMB primordial structure + radio sources + receiver noise and the null
evidence, which corresponds to a model without a cluster, i.e. simply
CMB + radio sources + receiver noise. The first model, which included
an SZ feature, was found to be $e^{15}$ times more probable than one without.

\begin{figure}
\centering
\includegraphics[width=7.5cm,height=
    7.5cm,clip=,angle=0.]{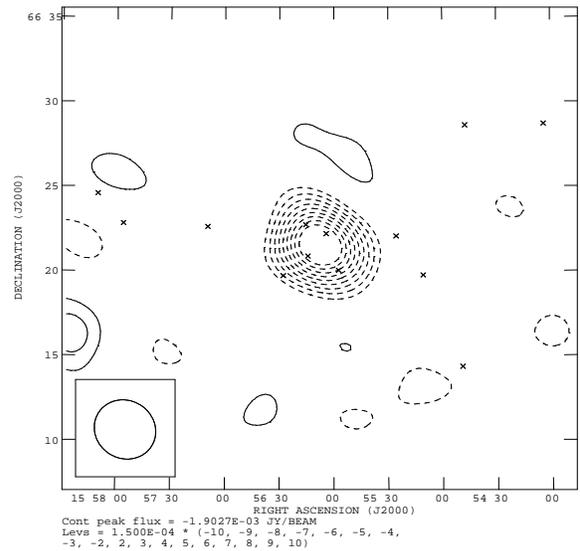}
\caption{ SA map after source subtraction using a 0.6-k$\lambda$
  taper. The crosses represent the position of the sources detected on the LA
map.}
\label{fig:SAmaftap}
\end{figure}

In Fig. \ref{fig:SAmaftap} a $0.6$-k$\lambda$ taper is used to
enhance large scale structure and consequently the signal-to-noise
ratio of the SZ effect. The peak decrement in it is $\approx 13 \sigma$.

The AMI SZ maps are compared to the Chandra X-ray emission and
projected temperature maps for Abell 2146 in the discussion, Section \ref{dis}.

\begin{table*}
\caption{List of the detected sources with their J2000 position
  coordinates, as determined by the LA map. Columns 3 and 4 show the
  flux densities of the detected sources
  at $16$-GHz ($S_{16}$) given by {\sc{McAdam}} using M3 with their
  associated Gaussian errors. For comparison, the LA measured flux densities at
  the same frequency are given. The letters represent the labeled sources in Fig. \ref{fig:SAmaf}.}
 
\label{tab:McAsourceFlux}
\begin{centering}
\begin{tabular}{|c|c|c|c|c|c|}
\hline 
{Source} & { RA (h m s)} & {Dec ($^{o}$ $^{'}$ $^{''}$)}
& {{\sc{McAdam}}-fitted $S_{16}$(mJy)} & {$\sigma$} & {LA $S_{16}$(mJy)}  \tabularnewline
\hline
\hline 
{ A} & { 15 56 04.23} & { +66 22 12.94} & { 5.92} & { 0.18} & {5.95}\tabularnewline
\hline 
{ B} & { 15 54 30.95} & { +66 36 39.58} & { 0.60} & { 0.29} & {0.61}\tabularnewline
\hline 
{ C} & { 15 56 14.30} & { +66 20 53.45} & { 1.83} & { 0.14} & {1.70}\tabularnewline
\hline 
{ D} & { 15 56 36.51} & { +66 35 21.65} & { 2.15} & { 0.15} & {1.65}\tabularnewline
\hline 
{ E} & { 15 55 57.42} & { +66 20 03.11} & { 1.65} & { 0.08} & {1.64}\tabularnewline
\hline 
{ F} & { 15 58 10.23} & { +66 24 35.72} & { 1.49} & { 0.12} & {1.29}\tabularnewline
\hline 
{ G} & { 15 54 03.96} & { +66 28 41.90} & { 1.12} & { 0.15} & {0.74}\tabularnewline
\hline 
{ H} & { 15 55 25.67} & { +66 22 03.96} & { 0.48} & { 0.05} & {0.67}\tabularnewline
\hline 
{ I} & { 15 55 10.84} & { +66 19 45.82} & { 0.61} & { 0.06} & {0.65}\tabularnewline
\hline 
{ J} & { 15 57 09.46} & { +66 22 37.62} & { 0.43} & { 0.06} & {0.63}\tabularnewline
\hline 
{ K} & { 15 54 47.50} & { +66 28 37.43} & { 0.91} & { 0.09} & {0.53}\tabularnewline
\hline 
{ L} & { 15 54 49.11} & { +66 14 21.49} & { 0.72} & { 0.09} & {0.47}\tabularnewline
\hline 
{ M} & { 15 56 15.40} & { +66 22 44.48} & { 0.16} & { 0.07} & {0.43}\tabularnewline
\hline 
{ N} & { 15 56 27.90} & { +66 19 43.82} & { 0.11} & { 0.05} & {0.33}\tabularnewline
\hline 
{ O} & { 15 57 56.10} & { +66 22 49.80} & { 0.30} & { 0.07} & {0.49}\tabularnewline
\hline
\end{tabular}
\end{centering}
\end{table*}

\subsection{Parameter estimates from three cluster models}\label{assumptions}

{\sc{McAdam}}  was run on the same Abell 2146 data for each of the
three models described in \ref{clusparam}. The results obtained for these
models are shown in Figs \ref{fig:2-D-m1} to \ref{fig:1-D-m4}. The contours in all the 2D marginalized posterior
distributions represent $68\%$
and $95\%$ confidence limits. Axis labels for
$M_{\rm{g}}(r_{200})$ are in units of $10^{13}$ for clarity.

\subsubsection{Cluster model 1}
\begin{figure}
\includegraphics[width=8.0cm,height=8.0cm,clip=,angle=0.]{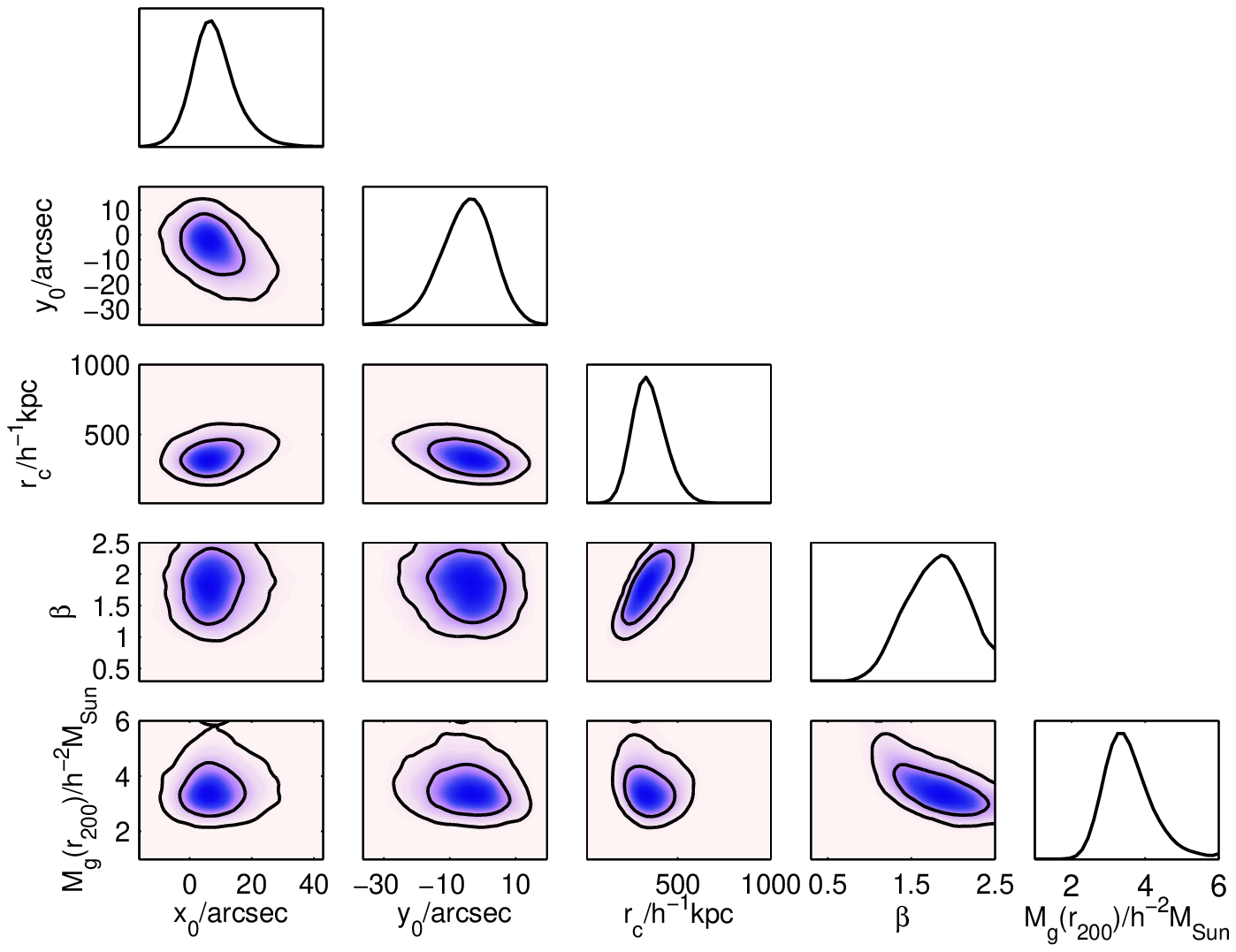}
\caption{Two-dimensional marginalized posterior distributions for the
  sampling parameters of Abell 2146---M1.}
\label{fig:2-D-m1}
\includegraphics[width=8.0cm,height=8.0cm,clip=,angle=0.]{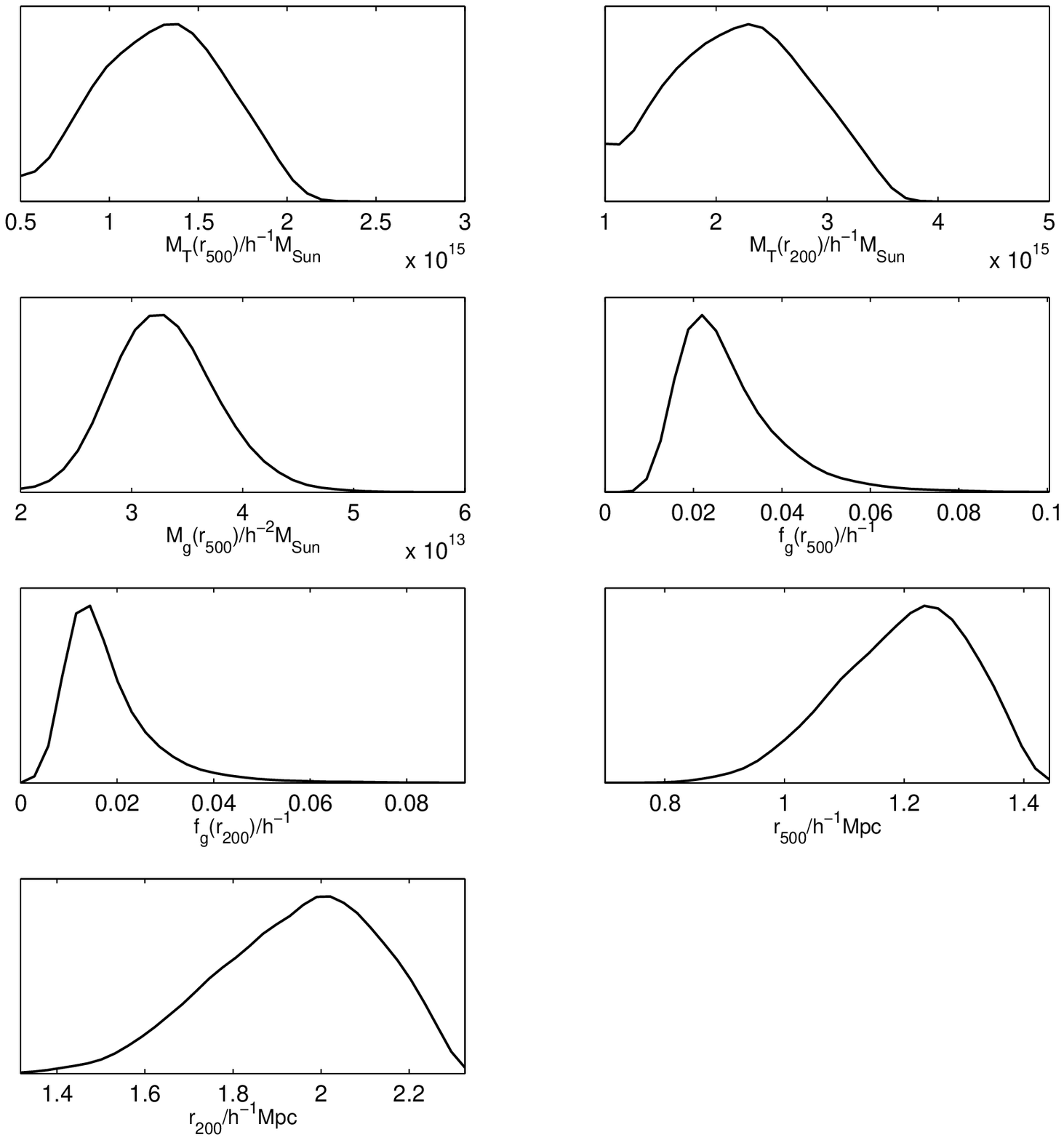}
\caption{One-dimensional posterior probability distributions for
  selected derived parameters of Abell 2146---M1. We note that the axes for
  the plots of the 1-D and 2-D marginalized posterior distributions
  of both the sampling and derived parameters are tailored to suit
  the results of each model and will therefore be different in each case}.
\label{fig:1-D-m1}
\end{figure}

The 2-D and 1-D marginalized posterior probability distributions for
the parameters of M1 are depicted in Figs. \ref{fig:2-D-m1} and \ref{fig:1-D-m1}, respectively.

M1 is representative of the more conventional method for extracting
cluster parameters from SZ data. In this model, the average cluster
gas temperature within $r_{200}$ is assumed to be known, from X-ray
measurements, allowing the morphology of the cluster,
namely $r_X$, to be inferred by assuming the cluster is spherical, in hydrostatic
equilibrium and described well by an isothermal $\beta$-model. The
overall bias on $r_X$ arises from all of these assumptions, which are
particularly unphysical in a cluster merger like Abell 2146, and is
therefore expected to be large. Indeed, by comparing Figs.
\ref{fig:1-D-m1} and \ref{fig:1-D-m4} we find that $r_{200}$ is
overestimated with respect to the value obtained in M3, our most
physically motivated model. Moreover, in M1, $M_{\rm{g}}(r_X)$ for $X=500$ and $1000$ depends on
$r_X$, which results in the bias on $r_X$ to be propagated to the
remaining derived parameters for these values of $X$.

\subsubsection{Cluster model 2}
\begin{figure}
\includegraphics[width=8.0cm,height=8.0cm,clip=,angle=0.]{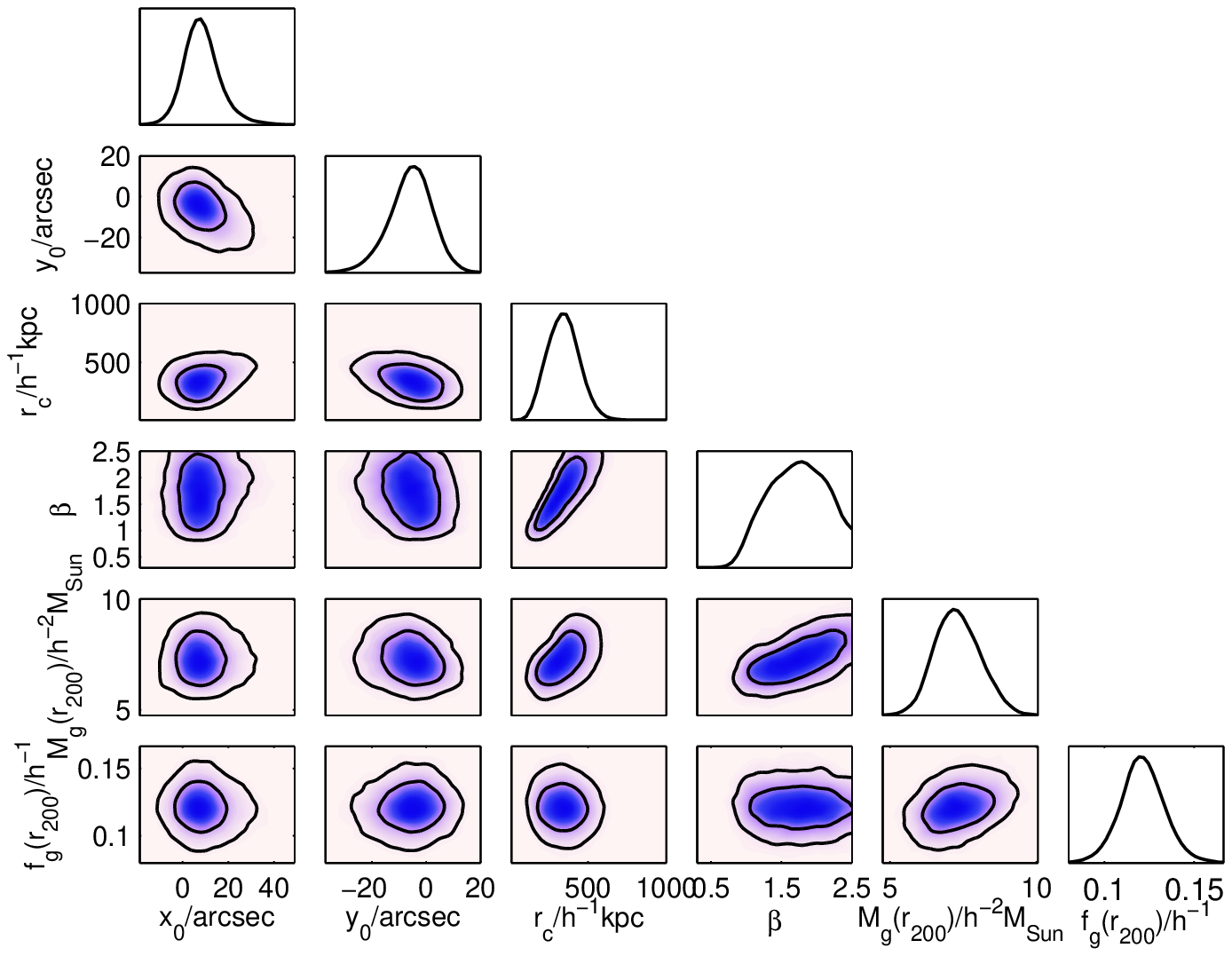}
\caption{Two-dimensional marginalized posterior distributions for the
  sampling parameters of Abell 2146---M2.}
\label{fig:2-D-m3}
\includegraphics[width=8.0cm,height=8.0cm,clip=,angle=0.]{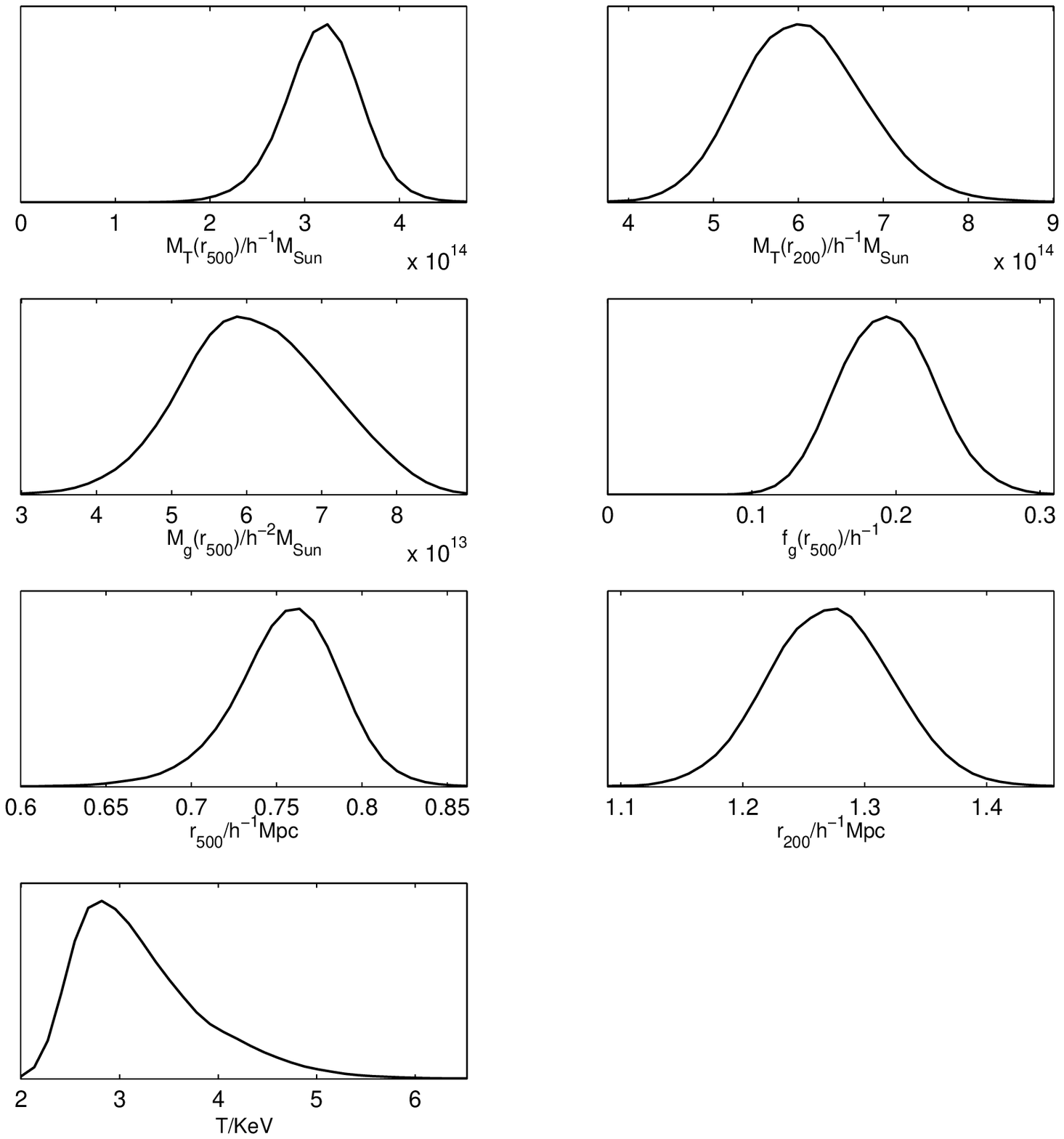}
\caption{One-dimensional marginalized posterior distributions for the
  derived parameters of Abell 2146---M2.}
\label{fig:1-D-m3}
\end{figure}

The 2-D and 1-D marginalized posterior probability distributions for
the parameters of M2 are depicted in Figs \ref{fig:2-D-m3} and \ref{fig:1-D-m3}.

M2 introduces a new sampling parameter, $f_{\rm{g}}(r_{200})$. Sampling from this
parameter allows more prior information to be included in the
analysis, which has the effect of constraining the parameter distributions
better than in M1. It has a great advantage over M1, namely, the only parameter obtained by
assuming hydrostatic equilibrium is the temperature, which is not used
explicitly in the calculation of the other derived parameters at $r_{200}$.

\begin{figure}

\includegraphics[width=7.5cm,height=
    7.5cm,clip=,angle=0.]{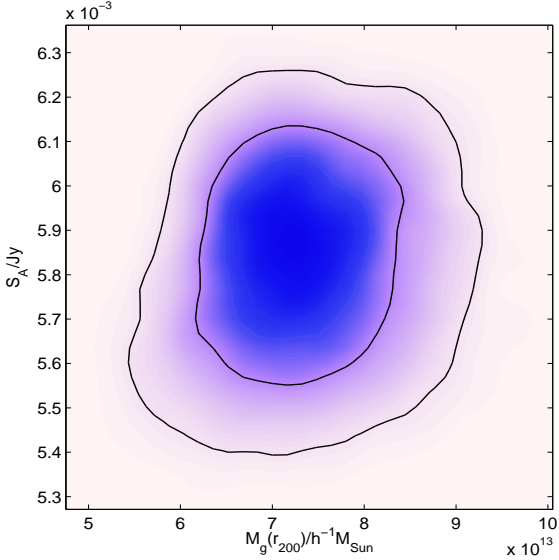}
\caption{Two-dimensional marginalized posterior distribution for the
  flux of source A shown in Fig. \ref{fig:SAmaf}, $S_A$, and the
  cluster gas mass within $r_{200}$, $M_{\rm{g}}(r_{200})$. }
\label{fig:massfluxrel}
\end{figure}

Fig. \ref{fig:massfluxrel} shows the two-dimensional marginalized
posterior distribution for the flux density of source A, $S_A$, and
$M_{\rm{g}}(r_{200})$--we choose to plot $S_A$ since source A is the
brightest source close to the pointing centre.
One can see from Fig. \ref{fig:massfluxrel} that $S_A$ and
$M_{\rm{g}}(r_{200})$ do not appear to be significantly
correlated. This is confirmed by the 
sample correlation, which was found to be 0.12. We note that the sample correlation remains unaffected
by shifts of origin or changes of scale in $S_A$ and 
$M_{\rm{g}}(r_{200})$. The flux density of source A is given a
Gaussian prior and yet the LA-measured and McAdam-derived flux-density
estimates for this source are very close. 

\subsubsection{Cluster model 3} \label{model4}
\begin{figure}
\includegraphics[width=8.0cm,height=8.0cm,clip=,angle=0.]{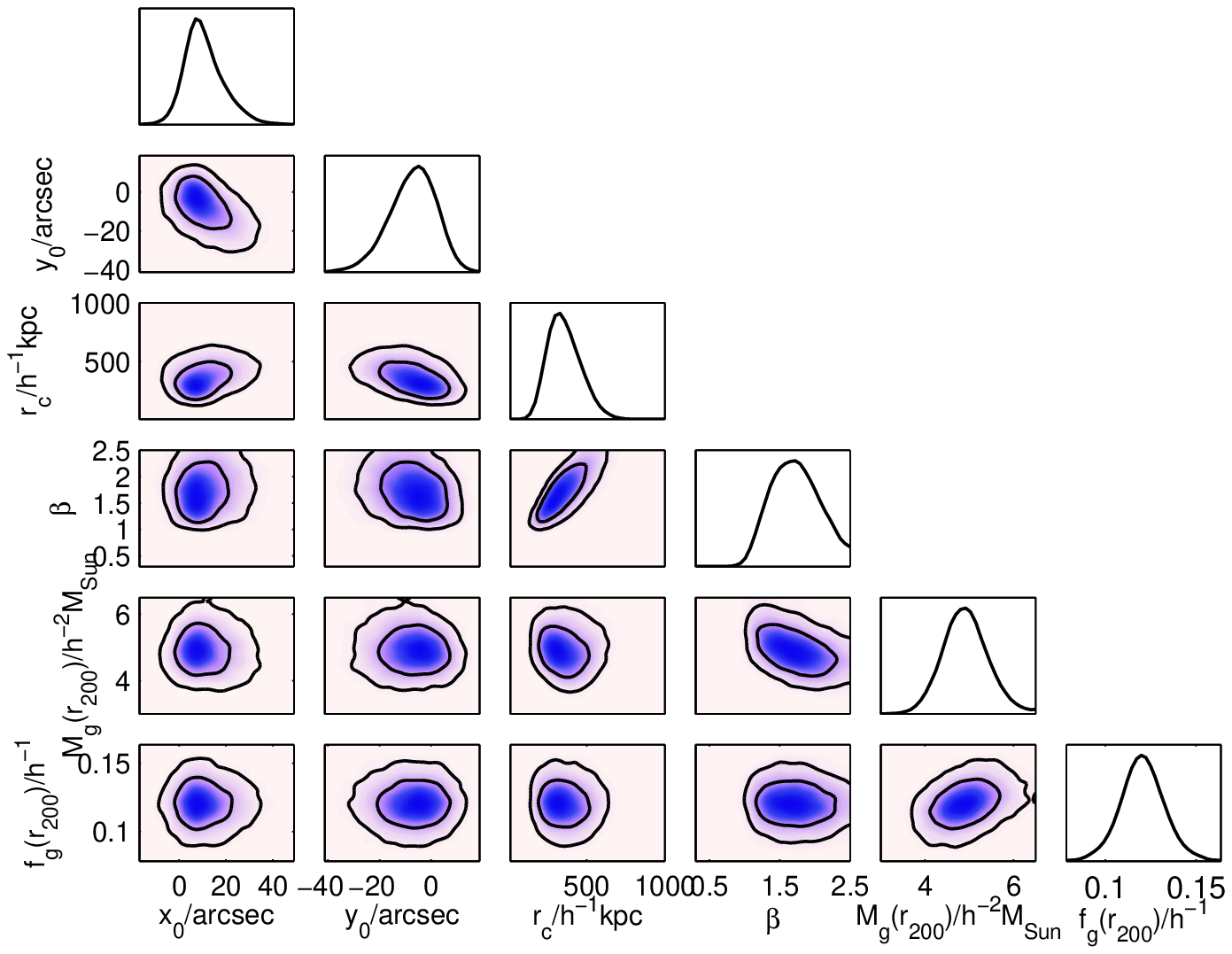}
\caption{Two-dimensional marginalized posterior distributions for the
  sampling parameters of Abell 2146---M3.}
\label{fig:2-D-m4}
\includegraphics[width=8.0cm,height=8.0cm,clip=,angle=0.]{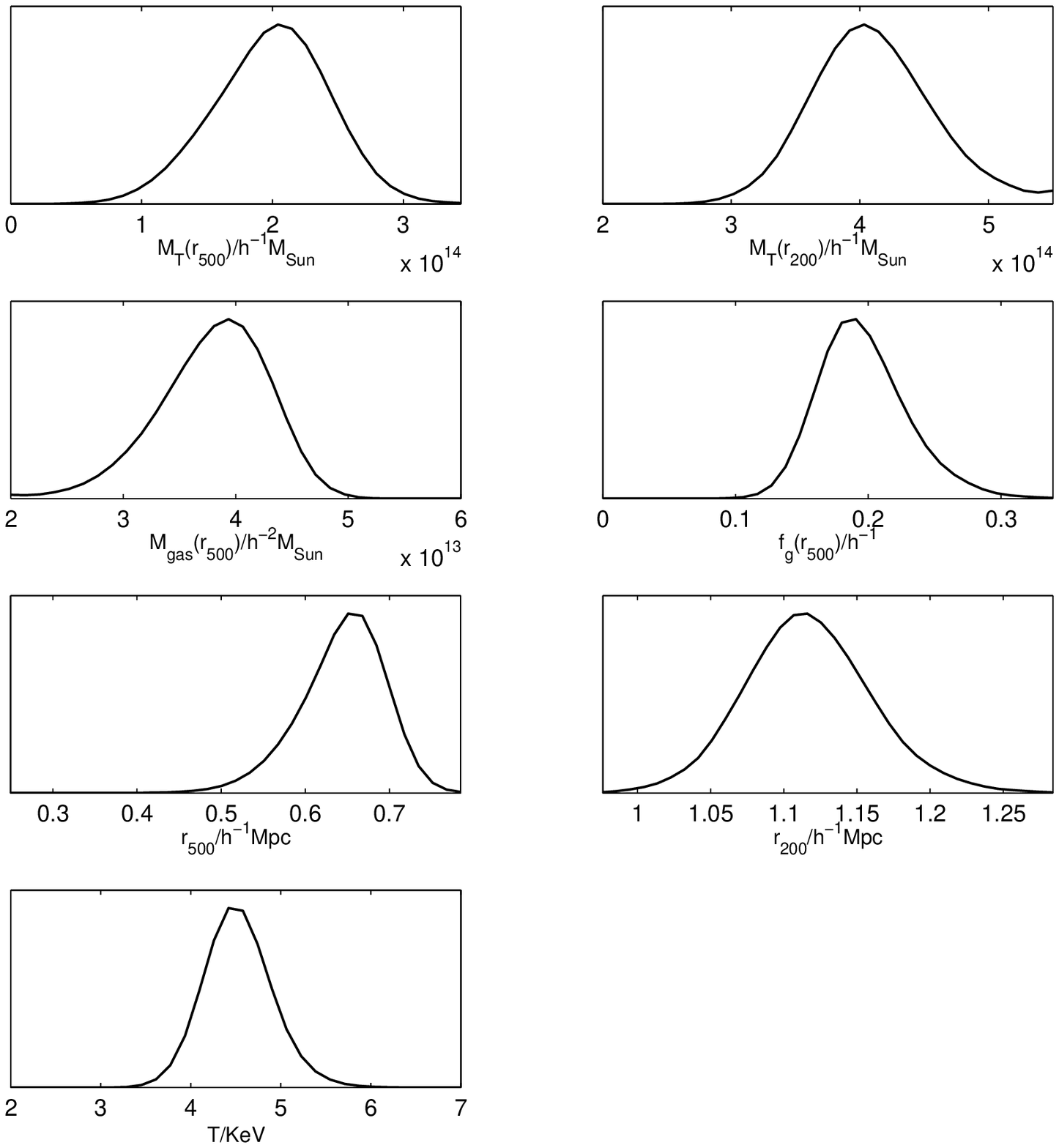}
 \caption{One-dimensional marginalized posterior distributions for the
 derived parameters of Abell 2146---M3.}
\label{fig:1-D-m4}
\end{figure}

The 1-D and 2-D marginalized posterior probability distributions for the parameters
of M3 are presented in Figs \ref{fig:2-D-m4} and
\ref{fig:1-D-m4}.

The only difference between M2 and M3 is in how the  average cluster
gas temperature at $r_{200}$, $T$, is calculated. To
obtain an estimate for $T$, M2 assumes the cluster is in hydrostatic
equilibrium while M3 uses the M-T relation in equation (\ref{eq:virtemp}), which assumes the
cluster is virialized and contains no unseen energy density.

\section{Discussion} \label{dis}
\subsection{Comparison with X-ray maps}
\begin{figure}
\centering
\includegraphics[width=7.5cm,height=
    7.5cm,clip=,angle=0.]{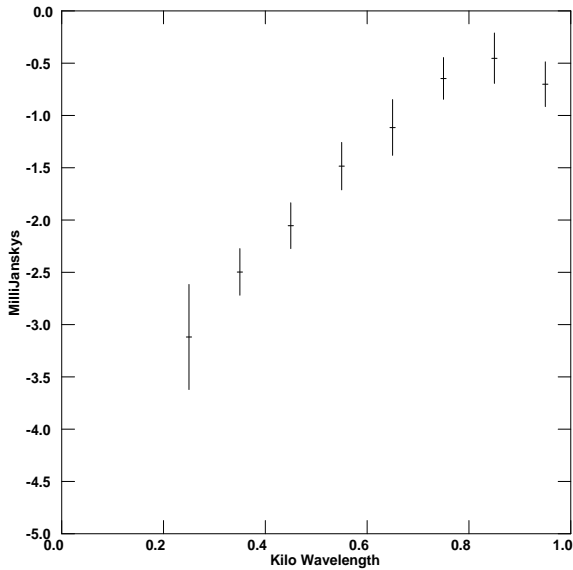}
\caption{Binned data from the {\sc{clean}}ed, radio source subtracted,
non-tapered map of Abell 2146 in bins of
$100\lambda$ against baseline (in k$\lambda$). It should be noted that
the FWHM of the aperture illumination function of the AMI SA is
$\approx185\lambda$ such that the visibilities in each bin are not
entirely independent. The baseline distance
corresponding to the {\sc{McAdam}} derived paremeters $r_{200}$,
$r_{500}$ and $r_{1000}$ were found to be
$0.42\rm{k}\lambda$,$0.816\rm{k}\lambda$, $1.46\rm{k}\lambda$, respectively.}
\label{fig:uvplt}
\end{figure}

\begin{figure}
\centering
\includegraphics[width=8.0cm,height=
    8.0cm,clip=,angle=0.]{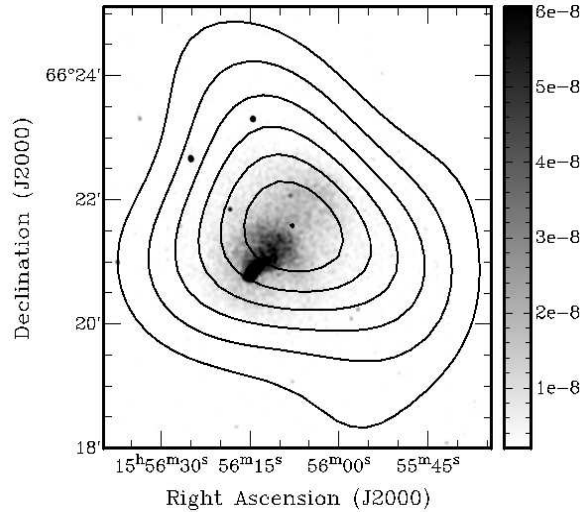}
\caption{Chandra X-ray image superimposed with AMI SA SZ effect (no taper).
The SA map is shown in black contours which go from $-1.4$ mJy$\,$beam$^{-1}$
to $0.001$ mJy$\,$beam$^{-1}$ in steps of $+0.2$ mJy$\,$beam$^{-1}$. The grey scale
shows the exposure-corrected image in the $0.3-5.0$ keV energy band
smoothed by a 2D Gaussian $\sigma=1.5$ arcsec (North is up and East is
to the left). The logarithmic scale bar has units of photons cm$^{-2}$s$^{-1}$arcsec$^{-2}$.}
\label{fig:xraySZ}
\end{figure}

Two new Chandra observations of Abell 2146 were taken in April
$2009$ \citep{russell09}. Fig. \ref{fig:xraySZ} shows the exposure-corrected
X-ray image taken in the $0.3-5.0$ keV energy band smoothed with
a 2D Gaussian of $\sigma=1.5$ arcseconds superimposed with the AMI SZ effect from Fig. \ref{fig:SAmaf}. The AMI $\it{uv}$-coverage is well-filled and goes down
to $\approx 180\lambda$ which corresponds to a maximum angular scale of
$\approx 10$ arcminutes or a cluster radius of $\approx
1.1$\:Mpc. Thus, in practice, the SZ signal traces a more extended region of the gas than the
X-ray data. Any small features in the cluster environment are not
resolved by the SA maps which consequently appear much more uniform than
the X-ray maps. Nevertheless, given the synthesized beams in
Figs. \ref{fig:SAmaf} and \ref{fig:SAmaftap}
the SZ effects in these two figures appear to show signs of some real
extended emission. To verify that we have resolved the SZ decrement
we bin the data from the {\sc{clean}}ed, radio source subtracted,
non-tapered map of Abell 2146, Fig. \ref{fig:SAmbef}, in bins of
$100\lambda$ and plot it against baseline, see
Fig. \ref{fig:uvplt}. The signal steadily becomes more negative from scales of
$800\lambda$ to $200\lambda$; it is on these larger scales that we find the most negative binned value for the
SZ decrement, demonstrating the sensitivity of the SA to large angular
scales. To determine the shape of the cluster in greater detail
high resolution SZ observations are needed.

During a cluster merger, elongations in the dark matter and gas
components are expected. In general, the orientation of this
elongation for both components tends to be parallel to the merger
axis, though the gas component can also be extended in a direction
perpendicular to the merger axis due to adiabatic compressions in the
ICM \citep{roettiger1997}, as shown in simulations of cluster mergers
\citep{poole2006}. We fitted a six-component (position, peak
intensity, major and minor axes and position angle) elliptical Gaussian to the SZ
decrement in our 0.6 tapered map,
Fig. \ref{fig:SAmaftap}, and a zero level using the {\sc{AIPS}} task {\sc{JMFIT}}. The results for the
parameters defining the shape of the fitted ellipse are given in Table
\ref{tab:jmfit}. The nominal results indicate that the semi-major
axis has a position angle of $46^{\circ}$. The orientation of the SZ
signal along this axis seems to be $\approx$orthogonal to the
elongation of the X-ray signal, see Fig. \ref{fig:xraySZ}.  Shock fronts
like the ones observed in Abell 2146 can only be detected during the early
stages of the merger, before they have reached the outer regions of
the system which suggests that the gas disturbances in the cluster
periphery are less intense than those near the dense core.

This is supported by the different signal distributions of
the X-ray and SZ effect data. The gas is relatively undisturbed in the
cluster periphery while in the inner regions the core passage has displaced the local gas at right angles to the merger axis \citep{russell09}. 

\begin{table}
\caption{{\sc{JMFIT}} results for the parameters of the ellipse fitted
  to the SZ decrement in the 0.6-tapered SA {\sc{clean}}ed maps. The extension of the minor and major
  axes are given in arcseconds and the position angle in degrees. }
\label{tab:jmfit}
\centering
\begin{centering}
\begin{tabular}{|c|c|c|c|} \hline
  {} & {Nominal} & {Minimum} & {Maximum} \\ \hline \hline
{Major axis} & {205} & {171} & {236} \\ \hline
{Minor axis} & {145} & {109} & {175} \\ \hline
{Position angle} & {46} & {3} & {68} \\ \hline
\hline 
\end{tabular}
\end{centering}
\end{table}

The total mass can also be estimated from the X-ray $M_{\rm{T}}(r_{500}) - T$
relation (eg. \cite{vikhlinin2006}) (note that here we use a different
scaling relation than elsewhere since we are concerned
with cluster parameters at $r_{500}$). Excluding the cool core region,
the X-ray spectroscopic temperature is $7.5\pm0.3$ keV, which
corresponds to a mass $M_{\rm{T}}(r_{500})\approx 7\pm2\times10^{14}
\rm{M}_{\odot}$ (using $h_{70}=1.0$).  This method will likely overestimate the cluster mass as
we expect the temperature to have been temporarily boosted during this major merger by a factor of a few
(\cite{ricker2001}, \cite{randall2002}).  A mass estimate for the Bullet
cluster from the $M_{\rm{T}}(r_{500}) - T$ relation produced a result approximately a factor of 2.4 higher than
the weak lensing result for the same region (\cite{markevitch2006}).  If we assume the X-ray mass estimate for Abell 2146 is overestimated
by a similar factor, the cluster mass should be closer to $M_{\rm{T}}(r_{500})\approx3\times10^{14} h^{-1} \rm{M}_{\odot}$, which is
comparable with our SZ effect result.  However, simulations show that
the transient increase in the X-ray temperature is dependent on the
time since the collision, the impact parameter of the merger and the
mass ratio of the merging clusters (e.g. \cite{ritchie2002}), which
will be different for the Bullet cluster.  A weak lensing analysis
using new Subaru Suprime-Cam observations will produce a
more accurate measure of the mass for comparison with the SZ effect result. 

\begin{figure}
\centering
\includegraphics[width=7.5cm,height= 7.5cm,clip=,angle=0.]{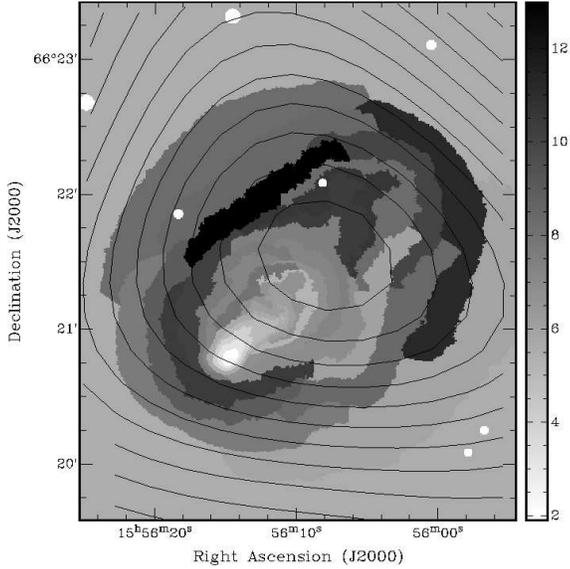}
\caption{Projected temperature map $\left(\textrm{keV}\right)$ \citep{russell09} overlaid
on black contours representing the SA SZ effect decrement. The contours go
from $-1.5$ mJy$\,$beam$^{-1}$ to $0.1$ mJy$\,$beam$^{-1}$ in steps of $+0.1$ mJy$\,$beam$^{-1}$
and the grey linear scale indicates the temperature variation
in keV.} 
\label{fig:ProjTemp}
\end{figure}

\subsection{Comparison with the 4.9-GHz VLA maps}

\begin{figure}
\centering
\includegraphics[width=7.5cm,height=
    7.5cm,clip=,angle=0.]{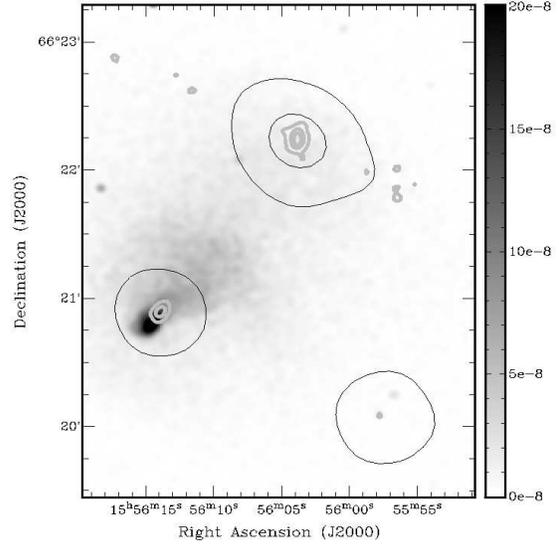}
\caption{VLA 4.9-GHz map in thick, grey contours overlaid on
the AMI LA map, in thin, black contours and the X-ray grey map from
Chandra observations. The logarithmic grey scale corresponds to the exposure-corrected
X-ray image taken in the $0.3-5.0$ keV energy band smoothed with
a 2D Gaussian of $\sigma=1.5$ arcseconds and it is in units of photons
cm$^{-2}$s$^{-1}$arcsec$^{-2}$. The VLA and LA contours range from  0.5 mJy$\,$beam$^{-1}$
to 9 mJy$\,$beam$^{-1}$ in steps of 0.3 mJy$\,$beam$^{-1}$.}
\label{fig:VLA-4.9-GHz}
\end{figure}

The VLA radio image taken at $4.9$-GHz (NRAO/VLA Archive Survey) and
the contours representing the LA map are superimposed on the X-ray
image in Fig. \ref{fig:VLA-4.9-GHz}. The presence of a bright source
on top of the dense cluster core obscures any possible high-resolution
SZ features in the LA map. High-resolution SZ images using the LA
would be possible if higher resolution data taken at 16-GHz were
available for source subtraction. The longer baselines of the LA
proved insufficient to remove the contaminant sources and no SZ effect
decrement was seen on the source subtracted LA maps. High-resolution
SZ effect measurements are necessary to disentangle the density and
temperature distributions properly. These observations in other
cluster mergers like the Bullet cluster \citep{malu2010} have revealed
structure in the gas pressure distribution and are powerful tools for
understanding the evolution of galaxy clusters.

Radio halos are faint, large-scale sources
that often span the entire cluster and are typically found in
cluster mergers. Two hours of VLA observations in two
configurations, C and D, towards A520 revealed a radio halo with a power of
$6.4\times10^{24}$W Hz$^{-1}$ \citep{govoni2001} at 1.4-GHz. The Bullet cluster
was also found to have a radio halo  with a power of
$(4.3\pm0.3)\times10^{25}$ W Hz$^{-1}$ at $1.3$-GHz \citep{liang2000}. No low
frequency radio data are currently available for Abell 2146. 4.9-GHz VLA
observations of Abell 2146 do not show signs for a radio halo, Fig. \ref{fig:VLA-4.9-GHz}, though deeper
observations, particularly at lower frequencies where radio halo emission
tends to be stronger, would be needed
to determine whether a radio halo is present in Abell 2146. Since
such halos are characterized by a steeply falling spectrum (e.g
\cite{hanisch1980, govono2004}) and no radio halo emission was
detected at 4.9-GHz, we do not
expect our observations to be contaminated by this diffuse emission. 

A520 and 1E0657-56 are the only two clusters that have been found to have both bow shocks and radio halos. They have 
provided unique information that allows determination of what proportion of the
ultrarelativistic electrons producing the radio halo are generated
as a result of merger-driven turbulence, as opposed to shock
acceleration \citep{markevitch2002, markevitch2006}. Since Abell 2146 is the
third cluster merger known to contain substantially supersonic shock fronts, finding a radio halo would
significantly improve our current understanding of how they are generated and
powered.

\subsection{Cluster Parameters}

The cluster parameters obtained from M3, our preferred model, are
discussed below.

\subsubsection{Position} \label{Position}
The mean value for the position,
$\rm{RA}\ 15^{\rm{h}}56^{\rm{m}}07^{\rm{s}}$ $\rm{Dec}\ +66^{\circ}21\farcm33\farcs$, with errors of
$6$ and $7$ arcseconds respectively, coincides with the X-ray centroid
position, $\rm{RA}\ 15^{\rm{h}}56^{\rm{m}}07^{\rm{s}}$ $\rm{Dec}\
+66^{\circ}21\farcm 35\farcs $, as
shown in Fig. \ref{fig:xraySZ}. However, the peak of the X-ray
flux is significantly displaced
from the peak of the SZ signal, as depicted in Fig. \ref{fig:xraySZ}.
The X-ray spectral luminosity is proportional to $\int n_{\rm{e}}^{2}T^{-1/2}\textrm{d}l$, while the SZ effect is a measure
of the integrated line-of-sight pressure and is proportional to $\int
n_{\rm{e}}T\textrm{d}l$. Therefore, the X-ray emission is more sensitive to substructure than
the SZ data and peaks at the position of the dense cluster core.

\subsubsection{$\beta$ and $r_{\rm{c}}$}

Results from running {\sc{McAdam}} on large samples of clusters have forced the prior on $\beta$ to be relaxed to include higher
values (see \cite{zwart2008}). The distributions for $\beta$ tend to favour higher values than
 typical X-ray estimates. However, this discrepancy is not surprising since previous
studies have revealed incompatibilities in the $\beta$ fits between
X-ray and SZ effect profiles due to their different dependencies
on parameters such as temperature and density \citep{hallman2007}.
The results show the degeneracy between $r_{\rm{c}}$ and $\beta$ but also
show evidence of strong constraints on this relation. This relation
is positively correlated in M1 and M2, where the assumption of
hydrostatic equilibrium is made to estimate parameters at $r_{200}$,
and negatively correlated in M3 where this assumption
is avoided.

\subsubsection{Gas fraction}\label{Gasf}

Sampling from $f_{\rm{g}}(r_{200})$ allows further prior information to be
introduced into the model which leads to better constrained parameter
estimates.

All the models were run through {\sc{McAdam}} without data to check the effect of the priors
on the results. From this test we discovered that in M1 the seemingly
inconspicuous priors on the sampling parameters lead to an effective
prior on $f_{\rm{g}}$ that peaks
around $0.01$ and strongly disfavours values of $f_{\rm{g}}\approx
0.1$. Since in our current models SZ data alone cannot place strong constraints on $f_{\rm{g}}$, the effective prior biases low
the estimates of $f_{\rm{g}}$ obtained in M1. On the other hand, when running M3 without data, the effective prior on
$f_{\rm{g}}$ does not change significantly from the Gaussian prior it was initially given. Given the
importance of analysing cluster models without any data to interpret
their results, a detailed discussion of these no-data runs and the
effects of cluster parameterization are presented in the
forthcoming paper, \cite{Malak2010}.

\subsubsection{Temperature}
The average cluster gas temperature within $r_{200}\approx 900$ kpc for
$h_{70}=1.0$ was found to be $4.5 \pm 0.5$
keV. The projected emission-weighted temperature map, Fig. \ref{fig:ProjTemp}, shows
a range of X-ray temperature measurements in different regions of the cluster.
At the position of the most negative value of the SZ decrement, the X-ray temperature
is $\approx8$ keV whereas at a radius of $\approx500$\:kpc the
temperature drops below 5 keV. In \cite{russell09}, a single-temperature fit to the cluster spectrum of Abell 2146 using an absorbed
thermal plasma emission model yields a temperature
of $6.7^{+0.3}_{-0.2}$\:keV. The higher X-ray temperature measurement is not surprising since
M3's derived temperature estimate refers to the mean cluster gas temperature
within $r_{200}$ and therefore averages over scales where the temperature
is lower. Moreover, emission-weighted temperatures will be higher
than mass weighted temperature estimates.

\subsubsection{Total mass}

Analytical and numerical simulations have already established
the integrated SZ signal as a robust tool for
determining the total cluster mass (see e.g. \cite{bart1994,
  barbosa1996, eke1996, dasilva2000, kravtsov2006, nagai2006, motl2005}). The measured SZ signal is sensitive to large scales away from
the cluster core and is therefore able to provide an estimate for
the $M_{\rm{T}}$ which is independent of the small-scale mechanisms that
regulate the state of the cluster gas near the core.

We find that, subject to the assumptions of M3 described in Section \ref{model3}, at the virial radius, $r_{200}$, 
$M_{\rm{T}}= \left(4.1 \pm 0.5 \right) \times 10^{14}
h^{-1}\rm{M}_\odot$; note that this estimate is free from the
assumption of hydrostatic equilibrium.

\section{Conclusion }\label{conc}

The AMI $16$-GHz observations of Abell 2146 presented in this paper show
the Sunyaev-Zel'dovich effect produced by this cluster with a peak 
signal-to-noise ratio of $13 \sigma$. We
detect 15 4-$\sigma_n$ sources within 0.1 of the primary beam in the SA pointed map using the
high resolution LA observations. These sources were subtracted from the
SA maps at the LA position using the flux densities obtained from running our
Bayesian analysis software, {\sc{McAdam}}, on the cluster model M3. Despite the substantial radio
emission from the direction of Abell 2146, no significant contamination from
radio sources is visible on the maps.

We compare our SZ observations with X-ray data taken by Chandra
and find an offset between the peaks of the two signals. We show that the
SA data resolves our SZ decrement and note that the directions of the
most pronounced elongations in the SZ and X-ray signals seem to be at $\approx 90^\circ$ to each other. These
results show complex dynamics indicative of a cluster merger and 
the differences in the gas emission and pressure distributions.

We run {\sc{McAdam}}, on three different cluster models, all of which
assume an isothermal, spherical $\beta$-model, and extract posterior
probability distributions of large-scale cluster parameters
of Abell 2146 in the presence of radio point sources, primordial CMB and
receiver noise. In M1, a model representative of more traditional
cluster parameterizations, the seemingly inconspicuous priors on the
sampling parameters lead to an effective prior on the derived
parameter $f_g(r_{200})$ which biases low this parameter and leads to further biases in other model parameters. 

M2 and M3 exploit the observation that the gas fractions do not appear to vary greatly
between clusters and sample directly from $f_{\rm{g}}(r_{200})$---introducing further constraints in our parameter space and avoiding
the bias problem in M1. The difference between M2 and M3 lies in the
derivation of the global cluster gas temperature, $T$. M2 assumes the
cluster is in hydrostatic equilibrium while in M3 $T$
can be deduced from the virial theorem (assuming all of the kinetic
energy is in the form of internal gas energy). Given the relative
masses of the two merging systems in Abell 2146 and, provided the primary cluster was virialized before the
merger, we find that the $T$ derived from the M-T
relation in M3 will change by $\approx 10\%$ K during the
merger. 

The results from M2 and M3 are consistent, despite differences in the
mean values of the large-scale cluster parameters. However, we choose to focus
on the results obtained in M3 since this model overcomes some of the
shortcomings of more traditional models and its global temperature estimate
is not significantly affected by the merger event. 
We find that at $r_{200}$ $M_{\rm{T}}= \left(4.1 \pm
0.5 \right) \times 10^{14} h^{-1}\rm{M}_\odot$,
$\beta=1.7 \pm 0.3$, $T=4.5 \pm 0.5$ keV and core radius $r_{\rm{c}}=358 \pm 100
 h^{-1}\,\textrm{kpc}$. We also find that the probability of SZ + CMB
primordial structure + radio sources + receiver noise to CMB + radio
sources + receiver noise is $3\times 10^{6}:1$.

\section*{Acknowledgments}\label{acknowledgements}
We thank the referee, Mark Birkinshaw, for helpful suggestions and comments.
We are grateful to the staff of the Cavendish Laboratory and the Mullard Radio Astronomy
Observatory for the maintenance and operation of AMI.
 We acknowledge support from the University of Cambridge and
 PPARC/STFC for funding and supporting AMI. ACF also acknowledges the
 Royal Society. CRG, HR, MLD, MO, MPS, TMOF, TWS  are grateful for
 support from PPARC/STFC studentships. This work was carried out using
 the Darwin Supercomputer of the University of Cambridge High
 Performance Computing Service (http://www.hpc.cam.ac.uk/), provided
 by Dell Inc. using Strategic Research Infrastructure Funding from the
 Higher Education Funding Council for England and the Altix 3700
 supercomputer at DAMTP, University of Cambridge supported by HEFCE
 and STFC. We thank Stuart Rankin for his computing support.


\end{document}